\documentclass[peerreview,draftcls,onecolumn,12pt]{IEEEtran}
\IEEEoverridecommandlockouts
\usepackage{cite}
\usepackage{amsmath,amssymb,amsfonts}
\usepackage{algorithmic}
\usepackage{graphicx}
\usepackage{textcomp}
\usepackage{xcolor}
\usepackage{comment}
\usepackage{textcomp}
\usepackage{array}
\usepackage{tabularx}
\usepackage{booktabs}
\usepackage{mathtools}
\usepackage{amsthm}
\usepackage{url}
\newcommand{\bigcell}[2]{\begin{tabular}{@{}#1@{}}#2\end{tabular}}

\DeclarePairedDelimiter\floor{\lfloor}{\rfloor}
\usepackage{subfigure}
\usepackage[font=small,labelfont=bf]{caption}
\usepackage{makecell}
\usepackage{pifont}
\newcommand{\cmark}{\ding{51}}%
\newcommand{\xmark}{\ding{55}}%
\newcommand{\makecellL}{\makecell[l]}%
\newcommand{\makecellC}{\makecell[c]}%

\usepackage{booktabs}
\def\BibTeX{{\rm B\kern-.05em{\sc i\kern-.025em b}\kern-.08em
    T\kern-.1667em\lower.7ex\hbox{E}\kern-.125emX}}
\theoremstyle{theorem}
\newtheorem{theorem}{Theorem}
\theoremstyle{definition}
\newtheorem{lemma}{Lemma}

\newtheorem{corollary}{Corollary}
\allowdisplaybreaks
\begin{document}
\title{Downlink Coverage and Rate Analysis of an Aerial User in Vertical Heterogeneous Networks (VHetNets)\\
\thanks{This work was supported in part by the Ministry  
of Higher Education and Scientific Research, Libya, through the Libyan-North American Scholarship Program, and in part by Huawei  Canada Co. Ltd.}
}
\author{\IEEEauthorblockN{  Nesrine~Cherif\IEEEauthorrefmark{1}, Mohamed Alzenad\IEEEauthorrefmark{2}, Halim Yanikomeroglu\IEEEauthorrefmark{2}, and Abbas Yongacoglu\IEEEauthorrefmark{1} \\
	\IEEEauthorblockA{\IEEEauthorrefmark{1}School of Electrical Engineering and Computer Science, University of Ottawa, Ottawa, ON, Canada\\
		\IEEEauthorrefmark{2}Department of Systems and Computer Engineering, Carleton University, Ottawa, ON, Canada	\\
		Email:	\IEEEauthorrefmark{1}\!\{ncher082, yongac\}@uottawa.ca,	\IEEEauthorrefmark{2}\!\{mohamed.alzenad, halim\}@sce.carleton.ca.}}

}
\maketitle
\vspace{-0.8cm}
\begin{abstract}
In this paper, we analyze the downlink coverage probability and rate of an aerial user in vertical HetNets (VHetNets) comprising aerial base stations (aerial-BSs) and terrestrial-BSs. The locations of terrestrial-BSs are modeled as an infinite 2-D Poisson Point Process (PPP), while the
locations of aerial-BSs are modeled as a finite 2-D Binomial Point Process (BPP). Our cellular-to-air (C2A) channel model incorporates line-of-sight (LoS) and non-LoS transmissions between terrestrial-BSs and a typical aerial user, while we assume LoS transmissions for all aerial links. We assume that  the aerial user is associated with an aerial-BS or terrestrial-BS that provides the strongest average received power. Using stochastic geometry, we derive exact and approximate expressions of the coverage probability and rate in terms of interference power's Laplace transform. The expressions are simplified assuming only LoS transmissions for the C2A channels.
This enables easy-to-compute equations with good accuracy at elevated aerial user heights. We find that aerial users hovering at low altitudes tend to connect to aerial-BSs in denser terrestrial environments. Employing directive beamforming at aerial-BSs guarantees an acceptable performance at the aerial user by reducing interference signals received from the aerial-BSs. In denser terrestrial networks, the performance at the aerial user degrades substantially despite beamforming.

\end{abstract}
\vspace{-0.8cm}
\vspace{-0.5cm}
\section{Introduction}
Unmanned aerial vehicles (UAVs, also referred to as drones) have drawn a great deal of interest in recent years due to their flexibility and cost effectiveness in executing a range of applications, including search-and-rescue missions, aerial imaging, surveillance, and package-delivery, to name few \cite{cao2018airborne}. To accommodate such high-capacity and demanding scenarios, ubiquitous coverage for UAVs is becoming crucial for delivering ultra-reliable and low-latency real-time communications \cite{zeng2019accessing,irem}. On this point, it may be worth mentioning that 3GPP in its recent report \cite{3gppTS22125} discussed the great potential of UAVs for providing cellular connectivity in 5G networks.  More specifically, the report investigated the communication needs of  UAVs (referred to as on-board radio access nodes (UxNB) in \cite{3gppTS22125}). Connected with this, a new paradigm has recently emerged where  existing cellular networks (e.g., Long Term Evolution (LTE) networks)  and future cellular networks (5G and beyond) can be re-utilized to provide cellular connectivity to UAVs (aerial users) \cite{Hayat}. Indeed, some field trials have been carried out to substantiate the feasibility of terrestrial networks for supporting aerial user communications \cite{mei}. The authors of \cite{imran2009comparison}, for instance, provided a comprehensive study of the potential of terrestrial base stations (terrestrial-BSs) and aerial base stations (aerial-BSs) in providing cellular coverage to aerial users based on 4G deployment. Besides the multitude of applications where UAVs are deployed as aerial users, UAVs can also be equipped with mounted antennas (i.e., as aerial-BSs) to provide cellular connectivity in case of disasters where the terrestrial network is in failure \cite{Zhou2018icc,tomic2012toward}. Also, significantly, aerial-BSs can be deployed to support  terrestrial network connectivity for  high-capacity demands, such as concerts and sporting events. Due to the nature of temporary spikes in throughput demand, deploying aerial-BSs is the fastest and most cost-effective solution, compared to terrestrial-BSs \cite{Kumbhar,Gupta,faraj2018,she2019,nesrineglobecom}. Aerial-BSs have also been investigated as a promising solution for providing pervasive cellular connectivity in  remote areas where the installation of conventional terrestrial network infrastructure is deemed challenging \cite{Hayajneh2018performance}. Clearly, the deployment of aerial users is imminent due to their  flexibility and the great improvements of their payload capacity and extended flight duration. However, the performance of aerial users in terms of coverage and achievable throughput have not received much attention in the literature on integrated aerial and terrestrial networks (i.e., vertical heterogeneous networks or VHetNets) \cite{nesrineglobecom,cherif2020optimal,alzenadcoverage,alzenad}. In this paper, we leverage tools from stochastic geometry to derive closed-form expressions for the downlink coverage and achievable rate of a typical aerial user in VHetNets under the assumptions of a 2-D infinite terrestrial network (terrestrial-BSs distributed over an infinite 2-D region with a certain density) and a 2-D finite aerial network. The results provide some interesting insights for the future deployment of aerial users/BSs.
\vspace{-0.4cm}
\subsection{Motivation and Related Work}

Due to the emerging applications of UAVs as aerial-BSs and aerial users, the performance of VHetNets involving aerial and terrestrial users have attracted a great deal of interest.
While the research community has extensively studied  the efficiency of deploying UAVs as aerial-BSs \cite{chetlur,alhouranilet,alzenad, shen}, the study of aerial users in terms of coverage probability and achievable rates has been lacking \cite{mei,geraci,azari2,azari}. For instance, the authors in \cite{mei} presented a non-orthogonal multiple access (NOMA) technique for uplink communications between aerial users and terrestrial-BSs. In \cite{challita},  the authors investigated interference-aware path planning for aerial users using a deep reinforcement learning algorithm.  Moreover, the work in \cite{mei} tackled the interference problem in aerial communications. A thorough comparison between the existing network infrastructure and futuristic deployment scenarios involving  massive multiple-input multiple-output (MIMO) systems on the performance of  aerial
users was studied in \cite{geraci}. While both \cite{mei} and \cite{geraci} focused on terrestrial-BSs as servers for aerial users, neither explored the opportunity of deploying  aerial-BSs for the cellular connectivity of aerial users.

Based on the tools of stochastic geometry, the authors in \cite{azari2} and \cite{azari} investigated the feasibility of serving aerial and terrestrial users with a terrestrial wireless network. The authors claimed that the line-of-sight (LoS) condition between aerial users and terrestrial-BSs degrades the signal-to-interference-and-noise ratio \textsf{SINR} at the terrestrial users due to the strong interference signals received from the LoS aerial users. The same authors also presented several tuning parameters to combat these strong interference signals. However, they did not consider the possibility of using  aerial-BSs capable of serving aerial users in the event of a congested terrestrial wireless network. 

While the Poisson point process (PPP) assumption has become the baseline for modeling the locations of terrestrial-BSs, it is not suitable for modeling aerial networks, especially those with a small number of aerial-BSs  deployed  over a limited area. Since the Binomial point process (BPP) was adopted to model ad hoc terrestrial networks with a given number of nodes distributed over a circular area \cite{chen2012,Torrieri2012}, it presents a reasonable model for aerial networks while also assuring the tractability of mathematical analysis. It is noteworthy that the deployment of BSs (terrestrial or aerial) is not random in practice. However, by assuming BPP aerial network, the mathematical analysis remains tractable. For instance, the work in \cite{chetlur} was the first to consider a BPP model for aerial networks. In this work, the links between terrestrial users and aerial-BSs were assumed to be in LoS condition (non-LoS (NLoS) links were not considered). Such an assumption may not be applicable to aerial-BSs operating at very low heights and/or in dense environments due to the high likelihood of  NLoS occurrences in these scenarios. In addition, the work in \cite{chetlur} limited the association of  terrestrial users to only aerial-BSs, and it did not consider the possibility of the terrestrial user being served by a nearby terrestrial-BS that would provide a better channel condition than that provided by an aerial-BS, if there were one. Similar to the authors in \cite{azari}, the authors in \cite{5gwf} investigated the opportunity of reusing existing terrestrial networks to serve aerial users. Their work suggested several interference mitigation techniques both in the uplink and downlink transmissions in order to maintain an acceptable performance of the terrestrial network for both terrestrial and aerial users. In \cite{amer2018caching}, the use of coordinated multi-point transmissions to provide seamless cellular connectivity for aerial users was studied. In the same work, clustered small cell BSs were considered for serving aerial users in content-caching architecture, where  popularly requested content was cached for aerial users which could then be transmitted to terrestrial users. However, the work was limited to scenarios where terrestrial-BSs did not coexist with aerial-BSs. The authors in \cite{alzenadcoverage} proposed a framework to analyze the coverage and rate of a  VHetNet. Yet the analysis was  exclusively for a typical terrestrial user; so it is not applicable for aerial users.

In Table \ref{tab:lit}, we summarize the stochastic geometry-based works in VHetNets in terms of characteristics for terrestrial and aerial networks, C2A transmissions link type, derived performance, and whether they account for aerial users. It may be worth mentioning that aerial-BSs can serve   terrestrial users along with their aerial counterparts. However, in this paper, we focus on filling the gap in the literature by investigating only the aerial user's performance in a VHetNet setup.
\begin{table}[!ht]
	\caption{The main relevant stochastic geometry-based works on VHetNets}
	\label{tab:lit}
\centering
	\begin{tabular}{l|l | l | l | l | l | l}
		\hline \hline
		\bfseries{\makecellL{Ref.\\ }}   & 
		\bfseries{\makecell[l]{Aerial \\ network}}  & \bfseries{\makecellL{Terrestrial \\ network}}&  \bfseries{\makecellL{C2A links\\ model}} & \bfseries{\makecellL{Coverage\\  prob.}}  &
		\bfseries{\makecellL{Rate}}  & \bfseries{\makecellL{Aerial\\ user}}    \\  \hline
		\cite{nesrineglobecom}  & \makecellL{Finite 2-D\\ BPP}     & \makecellL{Infinite 2-D\\ PPP} & \makecellL{LoS links} & \cmark &  \xmark & \cmark        \\ 	\hline
		\cite{alzenadcoverage}  &   \makecellL{Infinite 2-D\\ PPP}   & \makecellL{N/A} & \makecellL{LoS/NLoS \\links } & \cmark &  \cmark & \xmark        \\ \hline
		 \cite{alzenad}  & \makecellL{Infinite 2-D\\ PPP}     & \makecellL{Infinite 2-D\\ PPP} & \makecellL{LoS/NLoS\\ links } & \cmark &  \cmark & \xmark        \\	\hline
	  \cite{chetlur}  & \makecellL{Finite 2-D\\ BPP}     & \makecellL{N/A} & \makecellL{LoS links } & \cmark &  \xmark & \xmark        \\ 	\hline
	  	  \cite{azari2}  &   \makecellL{N/A}   & \makecellL{Infinite 2-D \\PPP} & \makecellL{LoS/NLoS\\ links} & \cmark &  \cmark & \cmark        \\ \hline	
	    \cite{azari}  &   \makecellL{N/A}   & \makecellL{Infinite 2-D \\PPP} & \makecellL{LoS/NLoS \\links } & \cmark &  \xmark & \cmark        \\ \hline	  
       \cite{amer2018caching}  &   \makecellL{N/A}   & \makecellL{Infinite 2-D\\ PPP} & \makecellL{LoS/NLoS \\links } & \cmark &  \cmark & \cmark        \\ \hline
	  \cite{Zhou2019}  &   \makecellL{Single \\aerial-BS}   & \makecellL{Single \\terrestrial-BS} & \makecellL{LoS/NLoS\\ links } & \cmark &  \xmark & \xmark        \\ \hline
	  
	  \makecellL{ This \\paper}  &   \makecellL{Finite 2-D\\ PPP}    & \makecellL{Infinite 2-D \\PPP} & \makecellL{LoS/NLoS \\links } & \cmark &  \cmark & \cmark        \\ 
	  \hline \hline
	\end{tabular}
\end{table}

\vspace{-0.5cm}
\subsection{Contributions}
The contributions of this paper can be summarized as follows:

\begin{itemize}
    \item We propose a novel framework to analyze the downlink coverage and rate of an aerial user served by a VHetNet comprising of aerial-BSs and terrestrial-BSs. The model consists of terrestrial-BSs that follow an infinite 2-D PPP and aerial-BSs that follow a finite 2-D BPP deployed at a particular height above the ground \cite{chetlur}. We model the cellular-to-air (C2A) links according to the general framework provided by the International Telecommunications Union (ITU) in its recommendation report \cite{itu},  which incorporates LoS and NLoS transmissions occurring according to a given probability. By contrast, the air-to-air (A2A) channels are assumed to be in LoS condition due to the absence of obstacles between the aerial user and aerial-BSs. In order to alleviate the strong interference signals received from the aerial-BSs, we incorporate directional beamforming at the aerial-BSs, where the interfering main beam of the aerial-BS  is directed toward the typical aerial user with a certain probability depending on its antenna beamwidth. Two spectrum sharing policies between aerial-BSs and terrestrial-BSs (referred to as orthogonal spectrum sharing (OSS) and non-OSS (N-OSS)) are considered. 
    \item Since the expression of the LoS probability of C2A links is  difficult to use, and the widely used  air-to-ground (A2G) model in \cite{alhouranilet} and \cite{alhourani1} was exclusively developed for
terrestrial users, we derive an easy-to-use compact exponential function as an approximation using curve fitting. We subsequently show that the obtained simple LoS probability matches perfectly with the ITU model in \cite{itu}.
    \item We derive exact and approximate expressions for the coverage probability and rate in terms of the Laplace transform of interference power, and we show that the derived expressions match the simulations perfectly.
    \item Under the assumption that there are only LoS C2A links (no NLoS transmissions), closed-form expressions for the aerial user association probabilities and the Laplace transform of the terrestrial interference power are derived.
    \item The mathematical analysis of the interference received from the aerial-BSs is challenging, especially because of the beamforming that directs the main beam of the aerial-BS toward the typical aerial user with a certain probability. This assumption results in two dependent tiers of aerial-BSs, where the number of aerial-BSs in each tier depends on the other tier. Based on this setup,  we derive an exact expression for the Laplace transform of the interference received from aerial-BSs in terms of the Meijer-G function. To the best of our knowledge, there are no existing studies that present a closed-form expression for the interference received from aerial-BSs. In fact, closed-form expressions are often presented as integral expressions \cite{chetlur}. In this case, the derived closed-form expression simplifies the evaluation of the expressions of the coverage probability and rate. 
\end{itemize}

This paper is organized as follows. The system model is presented in Section II. In Section III, we derive expressions for the association probabilities and the Laplace transforms of the aggregated interference powers. Also in Section III, we  derive some useful distance distributions that are  used in subsequent sections. Closed-form expressions for the coverage probability and the average achievable rate are derived in Section IV. Finally, we validate and compare the derived analytical expressions with Monte Carlo simulations and investigate the impact of several system parameters on the performance of the VHetNets.

\vspace{-0.5cm}

\section{System Model}

\begin{figure}[!t]
	\centering
	\subfigure[VHetNet with 3-D BPP aerial network: $N$ aerial-BSs are uniformly distributed within a finite cylinder with height $\rm H_C$ and  radius $r_{\rm D}$]{
		\includegraphics[scale=0.4]{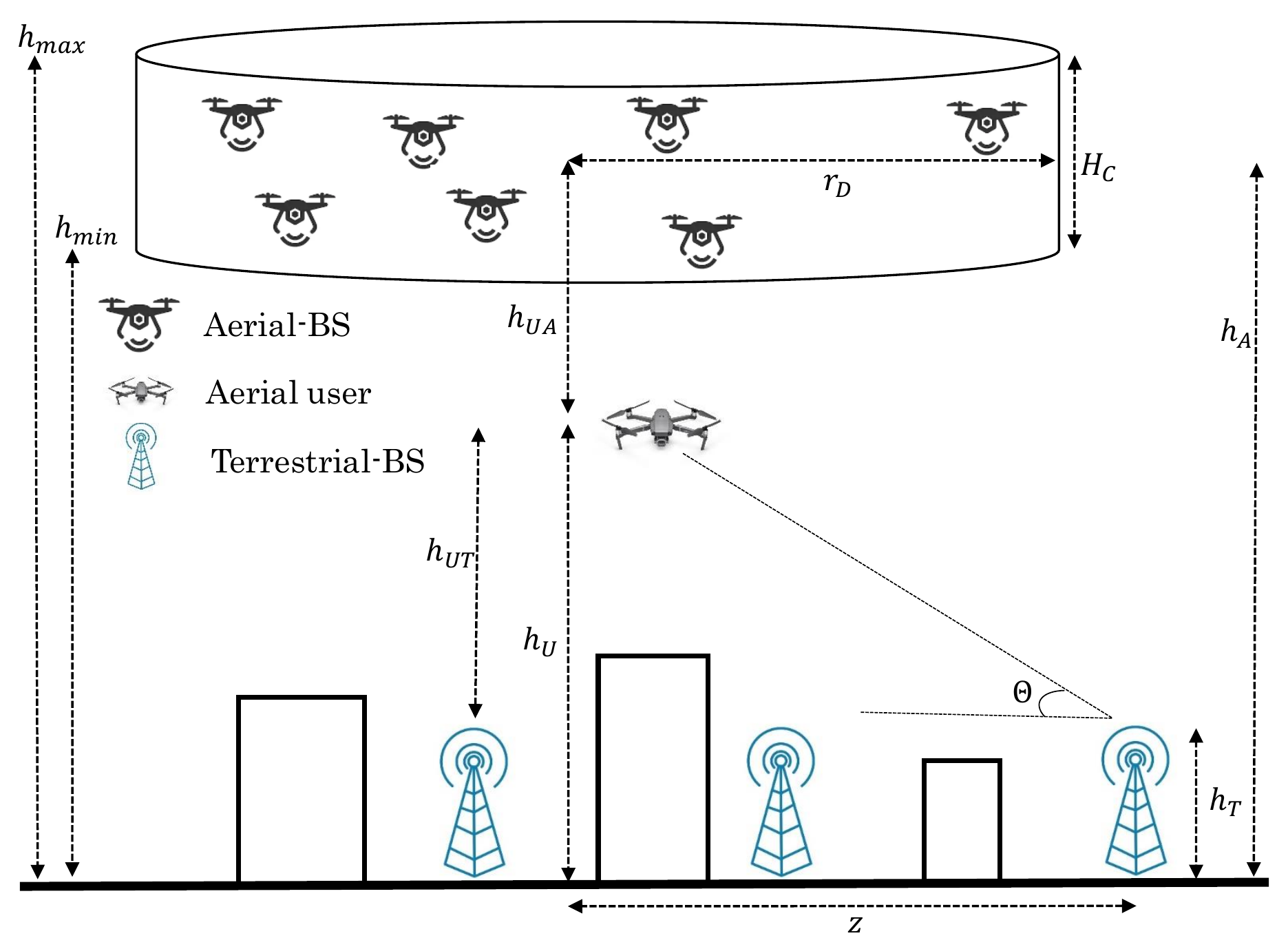}
	\label{fig:sys3D}
	}
	\hfill
	\subfigure[VHetNet with 2-D BPP aerial network: $N$ aerial-BSs are uniformly distributed over a finite disc with height $h_{\rm A}$ and  radius $r_{\rm D}$]{
		\includegraphics[scale=0.4]{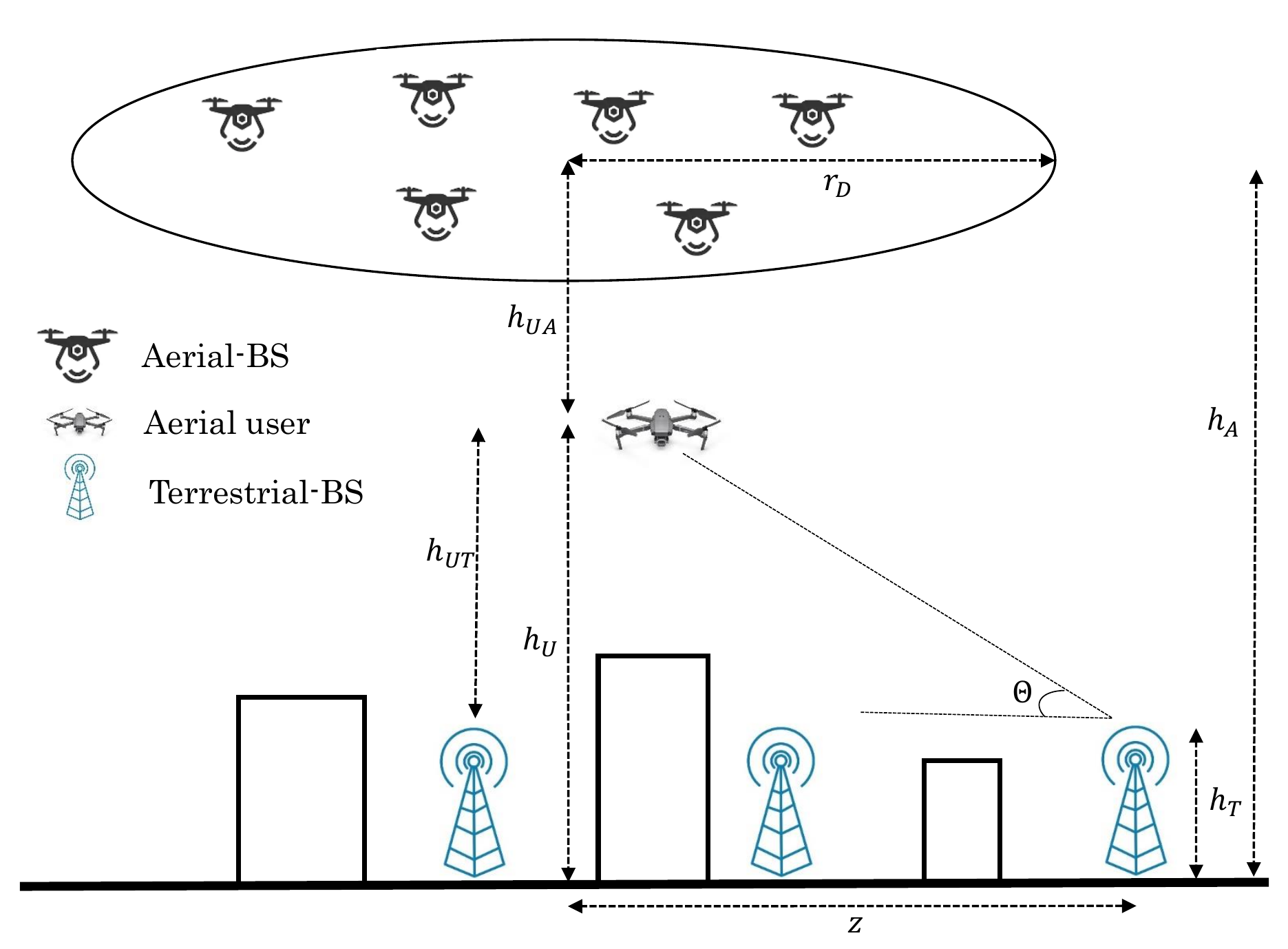}
		\label{fig:sys2D}
	}
	\caption[]{VHetNet with terrestrial-BSs, aerial-BSs, and aerial users.}
	\label{fig:3D2DBPP}
\end{figure}

We assume a network of single tier terrestrial-BSs, denoted by $\Phi_{\rm T}$, uniformly distributed at  ground level (i.e., terrestrial-BSs follow PPP) with density $\lambda_{\rm T}$ [terrestrial-BSs/km$^2$] and height $h_{\rm T}$ \footnote{Although terrestrial-BSs are often a mix of macro-, micro-, and pico-BSs, we only consider a single tier terrestrial network composed of macro-BS since macro-BSs transmit at higher powers than other terrestrial-BSs and are thus more suitable for providing wireless connectivity to aerial users.}. Since in practice, aerial-BSs can hover at different heights, we start by assuming that a finite number of aerial-BSs, $N$, hovers at different heights confined between  minimum and  maximum heights, $h_{\rm min}$ and $h_{\rm max}$, respectively, as shown in Fig. \ref{fig:sys3D}. We also assume that $N$ aerial-BSs are uniformly distributed within a cylinder with a radius $r_{\rm D}$ and a height $\rm H_C$ to form a BPP with a center $(0,0, \frac{h_{\rm min}+h_{\rm max}}{2})$. Introducing such a randomness to the height of aerial-BSs makes the distance distributions between a typical aerial user and any aerial-BS difficult to derive without much of a useful outcome. Since it has been shown in \cite{chetlur} that the performance of a 3-D BPP aerial network matches perfectly that of 2-D BPP aerial-BSs deployed at an average height, $ h_{\rm A}=\frac{h_{\rm min}+h_{\rm max}}{2}$, we assume in our framework, $N$ aerial-BSs hovering at a height $ h_{\rm A}$ from the ground and uniformly distributed over a disc  with a center $(0,0, h_{\rm A})$  and radius $r_{\rm D}$, as illustrated in Fig. \ref{fig:sys2D}. So although we place the aerial-BSs at the same height $ h_{\rm A}$, our framework is still applicable for aerial-BSs with different heights. Simulation results will substantiate the correctness of this assumption. It may be worth mentioning that the deployment of BSs (either terrestrial of aerial) is not random in practice. However, the point process assumption enables the tractability of the analysis of the network using stochastic geometry tools while tracking realistic BS deployment in practice \cite{andrews, alzenadcoverage, chetlur}. In this work, our focus is mainly on an aerial user's performance in VHetNets. The transmit power of the terrestrial-BSs and aerial-BSs are assumed to be $P_{T}$ and $P_{A}$, respectively. Finally, we consider a typical aerial user hovering at a height $h_{\rm U}$, where $h_{\rm T} < h_{\rm U}< h_{\rm A}$.
\vspace{-0.5cm}

\subsection{Cellular-to-air channel}
The widely used A2G channel model proposed in \cite{alhourani1} and \cite{alhouranilet} is not applicable to aerial users because it was developed for terrestrial users positioned at heights of 1.5 meters, while our system model involves terrestrial-BSs positioned at much greater heights (e.g., 30 meters for macro-BSs) \cite{baum2005interim}. Therefore, in this section, we develop a more tractable expression for the LoS probability.  According to the ITU recommendation report \cite{itu}, the probability of an LoS between a transmitter and receiver with heights $h_{\rm TX}$ and $h_{\rm RX}$, respectively, is given by \cite{itu}
\begin{equation}
\label{eq:itu}
P_{\rm L}(z)=\prod_{n=0}^{m}\left[1-\exp\left(-\frac{\left[h_{\rm TX}-\frac{\left(n+\frac{1}{2}\right)(h_{\rm TX}-h_{\rm RX})}{m+1}\right]^2}{2\delta^2}\right)\right],
\end{equation}
where $z$ denotes the horizontal distance between the transmitter and the receiver on the $xy$ plane, and $m=\floor{\left(\frac{z\sqrt{\alpha \beta}}{1000}-1\right)}$ where $\alpha$, $\beta$, and $\delta$ are environment-related coefficients given in Table I in \cite{holis}.

\begin{figure}[!t]
	\centering
	\includegraphics[scale=0.8]{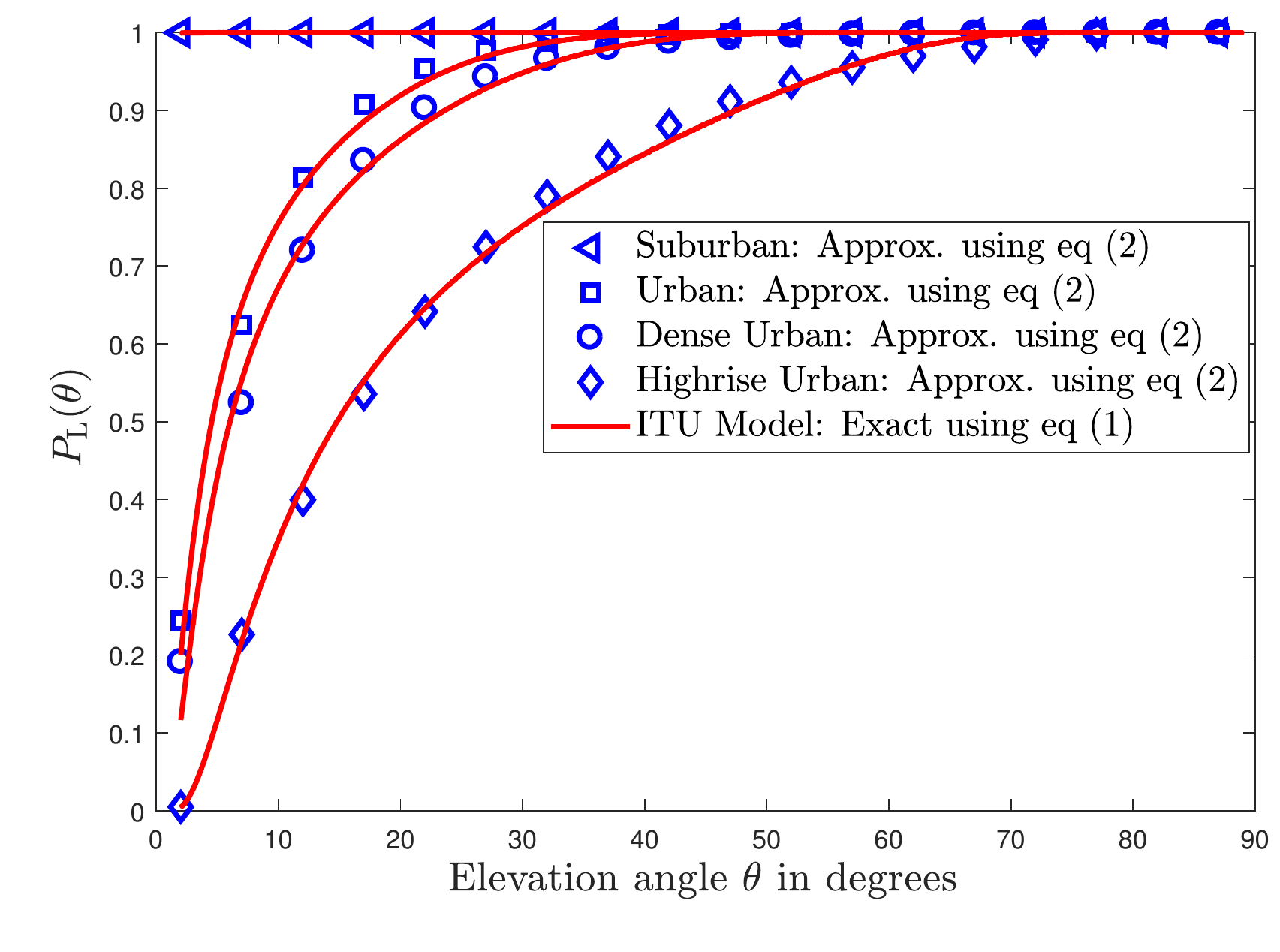}
	\caption{LoS probability versus the elevation angle $\theta$ for different environments.}
	\label{fig:itumodel}
\end{figure}

 As we can see in (\ref{eq:itu}), the LoS probability $P_{\rm L}(z)$ is not a continuous function of $z$ which makes the analysis intractable. Similar to the work in \cite{alhourani1} and \cite{alhouranilet}, we simplify (\ref{eq:itu}) to a more tractable  exponential function under some assumptions. This can be done by approximating the curve of $P_{\rm L}(z)$ in (\ref{eq:itu}) by a continuous exponential function of the elevation angle $\theta$ for a fixed terrestrial-BS height. From Fig. \ref{fig:sys2D}, the horizontal distance $z$ can be written as  $z=(h_{\rm U}-h_{\rm T})/\tan(\theta)$.  Comparing the parameters in (\ref{eq:itu}) with the parameters of the system model presented in Fig. \ref{fig:sys2D}, we have, $h_{\rm TX}=h_{\rm T}$ and $h_{\rm RX}=h_{\rm U}$. Notably, the resulting plot of the expression in (\ref{eq:itu}) will smooth for large values of $h_{\rm U}$ (we used $h_{\rm U}=10,000$  meters  in our simulation). Thus, the probability of an LoS condition becomes a continuous function of the elevation angle $\theta$ and the environment parameters. In Fig. \ref{fig:itumodel}, we plotted the ITU LoS probability for four different environments with a solid red line. As we can see, the trend of the curves can be closely approximated to a continuous exponential function. Using the Matlab\textsuperscript{\textregistered}  feature `Curve Fitting', we have
 \vspace{-0.5cm}
\begin{equation}
\label{eq:apprxitu}
P_{\rm L}(\theta)=-a\exp\left(-b \theta\right)+c,
\end{equation}
where $\theta=\tan^{-1}\left(\frac{h_{\rm U}-h_{\rm T}}{z}\right)$, and $a$, $b$, and $c$ are parameters that depend on the environment and height of the terrestrial-BSs. The LoS probability model used in this work is presented in Table \ref{my-label}. Indeed, the LoS probability derived in (\ref{eq:apprxitu}) for C2A links presents a compact expression that depends only on
the environment and the height of the aerial user in its elevation angle for a given terrestrial-BS
height.
\begin{table}[!t]
	\caption{Approximated LoS Probabilities for Different Environments}
	\label{my-label}
\centering
	\begin{tabular}{c c c c c}
		\hline \hline
		\bfseries{Environment}    & \bfseries $\mathbf{h_{\rm T}} $ (meters)  & \bfseries{a} & \bfseries{b} & \bfseries{c}  \\  \hline
		Suburban   & 30     & 1 & 6.581 & 1          \\ 
	   	Urban   & 19      &    1  &    0.151 &   1   \\ 
	   	Dense urban   & 25       &  1 &   0.106 &   1  \\
	   	Highrise urban   & 62      &  1.124 &  0.049 &  1.024     \\ \hline \hline
	\end{tabular}
\end{table}

Fig.~\ref{fig:itumodel} shows the LoS probability using the original ITU model in (\ref{eq:itu}) and the approximated expression in (\ref{eq:apprxitu}) for different environments. As we can see, the approximated LoS expression in (\ref{eq:apprxitu})  matches the original ITU model very closely. Each C2A link between the aerial user, which is at a height $h_{\rm UT}=h_{\rm U}-h_{\rm T}$ from the terrestrial-BS, is assumed to be either an LoS or  NLoS link with an LoS probability $P_{\rm L}(z)$ given by 
\begin{equation}\label{eq:PL}
P_{\rm L}(z)=-a\exp\left(-b \tan^{-1}\left(\frac{h_{\rm UT}}{z}\right)\right)+c.\\
\end{equation} 
Finally, the NLoS probability is given by $P_{\rm N}(z)=1-P_{\rm L}(z)$.

Due to the fact that the aerial user is either in an LoS or NLoS condition with each terrestrial-BS \cite{alzenad}, the set of terrestrial-BSs can be broken down into two independent inhomogeneous PPPs, where the LoS terrestrial-BSs form a subset $\Phi_{\rm L}$ with density $\lambda_{\rm T} P_{\rm L}(z)$, while the NLoS terrestrial-BSs form a subset $\Phi_{\rm N}$ with density $\lambda_{\rm T} P_{\rm N}(z)$ \cite{haenggi2012stochastic}. We assume that the LoS and NLoS C2A channels experience Nakagami-$m$ fading with different $m$ parameters, and therefore the received power from the terrestrial-BS located at point $x_j$ is Gamma-distributed, i.e., ${\rm H}_{\nu}^{x_j}\sim \text{Gamma}\left(m_{\nu},\frac{1}{m_{\nu}}\right)$, where $m_{\nu}$, $\nu \in \{\rm L,N\}$ denotes the fading parameters for the LoS and NLoS C2A links. The probability density function (PDF) of the Gamma-distributed channel gain is given by \cite{alzenad}
\vspace{-0.3cm}
\begin{equation}
\label{eq:PDFGa}
f_{{\rm H}_{\nu}^{x_j}}(x)=\frac{m_{\nu}^{m_{\nu}}x^{m_{\nu} x}}{\Gamma({m_{\nu}})}e^{-m_{\nu} x}, \quad \nu \in \{\rm L,N\},
\end{equation}
where $\Gamma(\cdot)$ is the Gamma function defined as $\Gamma(m_{\nu})=\int_{0}^{\infty} t^{m_{\nu}-1} e^{-t} dt$.

The received power at the typical aerial user from a terrestrial-BS located at $x_j$ is given by
\vspace{-0.3cm}
\begin{equation}
\label{eq:PrTBS}
P_{r,j}^{ \nu}= P_T \eta_{\nu} G^{\rm T}(r_j){\rm H}_{\nu}^{x_j}d_{\nu,x_j}^{-\alpha_{\nu}}, \quad \nu \in \{\rm L,N\},
\end{equation}
where $\eta_{\nu}$ are the excess losses for the LoS and NLoS C2A links, respectively 
\cite{alhourani1}. Moreover, $d_{\nu,x_j}$ denotes the distance between the typical aerial user and a terrestrial-BS from  the tier $\Phi_{\nu}, \nu \in \{\rm L,N\}$ located at $x_j$.  $\alpha_{\nu}$, $\nu \in \{\rm L,N\}$, is the path-loss coefficient. Finally, $G^{\rm T}(r_j)$ denotes the antenna directivity gain of the terrestrial-BS's antenna located at $x_j$ and the typical aerial user, which can be written as \cite{azari}
\begin{equation}
\label{eq:gainTBS}
G^{\rm T}(r_j)=\left\{
\begin{array}{ll}
G_{\rm m}^{\rm T} ,& r_j \in \mathcal{D}_{\rm TBS}\\
g_{\rm s}^{\rm T}, & r_j \notin \mathcal{D}_{\rm TBS},
\end{array}
\right.
\end{equation}
where $\mathcal{D}_{\rm TBS}$ denotes the set of all distances $r_j$ satisfying $r_j\tan \left(\theta_t+\frac{\theta_{\rm B}^{\rm T}}{2}\right)<h_{\rm U}<h_{\rm T}-r_j\tan \left(\theta_t-\frac{\theta_{\rm B}^{\rm T}}{2}\right)$, where $r_j$ refers to the horizontal distance separating the aerial user from the projection of the terrestrial-BS location on the $xy$ plane, and $\theta_t$ and $\theta_{\rm B}^{\rm T}$ are the down-tilt and beamwidth angles of the terrestrial-BS's antenna. In other words, $G^{\rm T}(r_j)$ determines whether the aerial user falls within the mainlobe or sidelobe of the terrestrial-BS's antenna.

Since the terrestrial-BS antennas are typically tilted toward the ground in order to provide cellular coverage to terrestrial users \cite{3gpp36814}, aerial users that hover at a greater height than the terrestrial-BS, i.e., $h_{\rm T}< h_{\rm U}$, will  most likely receive signals from the terrestrial-BS antenna sidelobes. For the tractability of the analysis in this paper, we assume that $G^{\rm T}(r_j)=g_{\rm s}^{\rm T},\ \forall\ r_j$. In the simulation results section, we will provide findings to substantiate the accuracy of this assumption. 
\vspace{-0.5cm}

\subsection{Air-to-air channel}
Since the aerial user is assumed to hover above rooftops, the A2A links between the typical aerial user and the aerial-BSs  are assumed to be in LoS condition \cite{walter2010uhf,goddemeier2015investigation}. We also assume that the aerial-BSs employ directive beamforming to improve the \textsf{SINR}. As a result, the gain of the aerial-BS antenna located at $x_i$, denoted by $G^{\rm A}$, at the typical aerial user is given by \cite{mozaffari2016efficient}
\begin{equation}
\label{eq:gain}
G^{\rm A}(d_{A,x_i}) =\left\{
\begin{array}{ll}
G_{\rm m}^{\rm A} ,& -\frac{\theta_{\rm B}^{\rm A}}{2}\leqslant\psi\leqslant\frac{\theta_{\rm B}^{\rm A}}{2}\\
g_{\rm s}^{\rm A}, &\text{otherwise},
\end{array}
\right.
\end{equation}
where $d_{A,x_i}$ denotes the distance between the typical aerial user and the aerial-BS located at $x_i$, $G_{\rm m}^{\rm A}$ and $g_{\rm s}^{\rm A}$ are the gains of the mainlobe and sidelobe, respectively, $\psi$ is the sector angle, and $\theta_{\rm B}^{\rm A}\in[0,180]$ is the beamwidth in degrees \cite[Fig. 3]{chetlur2019coverage}. Thus, the received power at the aerial user from  an aerial-BS located at $x_i$ is given by
\vspace{-0.3cm}
\begin{equation}
P_{r,i}^{ A}= P_A G^{\rm A}(d_{A,x_i})\eta_{A} {\rm H}_{\rm A}^{x_i}d_{A,x_i}^{-\alpha_{A}},
\end{equation}
where  $\eta_{A}$ represents the excess losses, ${\rm H}_{\rm A}^{x_i}$ is the Gamma-distributed channel power gain, i.e., ${\rm H}_{\rm A}^{x_i}\sim\text{Gamma}\left(m_A,\frac{1}{m_A}\right)$, with a fading parameter $m_A$, and $\alpha_{A}$ is the path-loss exponent. Unlike terrestrial-BSs whose antennas with fixed antenna patterns are deployed to serve terrestrial users, the aerial-BSs with dwon-tilted antennas can direct their beams toward their associated aerial users. Hence, the beamforming gain from the serving aerial-BS is always $G_{\rm m}^{\rm A}$ \cite{chetlur2019coverage}. For the remaining interfering aerial-BSs, their main beams are not necessarily aligned with the typical aerial user since their beam orientations are uniformly distributed in $[0,180^{\circ}]$. We introduce a probability $q_{\rm A}$ depending on the beamwidth $\theta_{\rm B}^{\rm A}$ that quantifies the likelihood of an interfering aerial-BS's mainlobe being directed toward the typical aerial user. It follows that $q_{\rm A}=\mathbb{P}\left(G^{\rm A}(d_{A,x_i})=G_{\rm m}^{\rm A}\right)=\frac{\theta_{\rm B}^{\rm A}}{180}$. The interfering aerial-BS gain is $g^{\rm A}_{\rm s}$ when the typical receiver is in its sidelobe direction with a probability of occurrence of $1-q_{\rm A}$ \cite[Fig. 3]{chetlur2019coverage}. Typically, $g^{\rm A}_{\rm s}$ is $20$ dB less than the mainlobe gain $G_{\rm m}^{\rm A}$ \cite{chetlur2019coverage}. It is worth noting that realistic 3-D antenna radiation models, such as those presented in \cite{haneyaPimrc} and \cite{haneyaTrans}, can be adopted in our VHetNet’s framework analysis. However, as shown above, we choose to investigate a more analytically tractable BS antenna model. Future works can focus on  more realistic 3-D antenna patterns in the context of VHetNets for even more practical insights.

The typical aerial user is assumed to be served by the BS (located at point $x_0$) that provides  the strongest \textit{long-term averaged} received power \cite{jo2012het}. Indeed, since the A2A links are in LoS, it is possible for a distantly located aerial-BS to offer better \textsf{SINR} than that of a closer terrestrial-BS due to differences in path-loss parameters. It should be noted that if the aerial user is associated with a specific tier of BS, i.e., $\{\Phi_{\rm L},\Phi_{\rm N},\Phi_{\rm A}\}$, the serving BS would be the nearest BS to that specific tier.  Thus, by assuming that the average power of all channels is 1, i.e.,  $\mathbb{E}[{\rm H}_{\rm L}^{x_j}]=\mathbb{E}[{\rm H}_{\rm N}^{x_j}]=\mathbb{E}[{\rm H}_{\rm A}^{x_j}]=1$ for each $x_j \in \Phi_{\rm T}  \cup\Phi_{\rm A}$, the serving BS is given by
\vspace{-0.3cm}
\begin{equation}
\label{eq:x0}
x_0=\text{arg} \max\{\mu_LR_{L}^{-\alpha_{L}}\!\!,\mu_NR_{N}^{-\alpha_{N}}\!\!,\mu_A R_{A}^{-\alpha_{A}}\},
\end{equation}
where  $\mu_{\nu}=P_T \eta_{\nu} g_{\rm s}^{\rm T}$, $\nu \in\{\rm L,N\}$, $\mu_A=P_A G_{\rm m}^{\rm A}\eta_{A}$, $R_L=\underset{\forall  x_j \in \Phi_{\rm L}}{\min} d_{L,x_j}$, and, $R_N=\underset{\forall x_j \in \Phi_{\rm L}}{\min} d_{N,x_j}$, and $R_A=\underset{\forall x_i \in \Phi_{\rm A}}{\min} d_{A,x_i}$. As a result, the \textsf{SINR} at the typical aerial user is given by
\begin{equation}
\label{eq:sinr}
\gamma=\left\{
\begin{array}{ll}
	\frac{\mu_L {\rm H}_{\rm L}^{x_0}R_{L}^{-\alpha_{L}}}{I+\sigma^2}, &{\rm if} \hspace{4pt} x_0 \in \Phi_{\rm L} (\mathcal{E}_{\rm L})\\
	\frac{\mu_N {\rm H}_{\rm N}^{x_0}R_{N}^{-\alpha_{N}}}{I+\sigma^2}, &{\rm if} \hspace{4pt} x_0 \in \Phi_{\rm N} (\mathcal{E}_{\rm N})\\
	\frac{\mu_A {\rm H}_{\rm A}^{x_0} R_{A}^{-\alpha_{A}}}{I+\sigma^2},  &{\rm if} \hspace{4pt} x_0 \in \Phi_{\rm A} (\mathcal{E}_{\rm A}),
\end{array}
\right.
\end{equation} 
 where  $\mathcal{E}_{\nu}$, $\nu \in \{\rm L,N,A\}$, denotes the event that the serving BS belongs to the tier $\Phi_\nu$ and $\sigma^2$ is the additive white Gaussian noise power. Finally, $I$ refers to the aggregate interference power. 
 
 Two spectrum sharing policies between aerial-BSs and terrestrial-BSs are used: orthogonal spectrum sharing (OSS) and non-orthogonal spectrum sharing (N-OSS) \cite{Zhou2019}. OSS/N-OSS implies that aerial-BSs and terrestrial-BSs operate on the same/different frequencies. The interference $I$ is given by 

\begin{equation}
\label{eq:int}
I=\left\{
\begin{array}{ll}
I_L+I_N+I_A, & \text{N-OSS}\\
I_L+I_N, &\text{OSS}, \hspace{8pt} \text{ if } x_0 \in \Phi_{\rm L}\cup \Phi_{\rm N}\\
I_A,  &\text{OSS}, \hspace{8pt} \text{ if }x_0 \in \Phi_{\rm A},
\end{array}
\right.
\end{equation} 
where
\begin{eqnarray}
\!\!\!\!\!\!\!\!&&I_L=\!\!\!\!\!\!\!\!\sum_{x_j\in\Phi_L \setminus x_0}^{}\!\!\!\!\mu_L {\rm H}_{\rm L}^{x_j} d_{L,x_j}^{-\alpha_L}, \hspace{6pt}I_N=\!\!\!\!\!\!\!\!\sum_{x_j\in\Phi_N \setminus x_0}^{}\!\!\!\!\mu_N {\rm H}_{\rm N}^{x_j} d_{N,x_j}^{-\alpha_N},\text{ and,  }I_A=\!\!\!\!\!\!\!\!\sum_{i=1,x_i\in\Phi_A \setminus x_0}^{N}\!\!\!\!\!\!\!\!P_A G^{\rm A}(d_{A,x_i})\eta_{A} {\rm H}_{\rm A}^{x_i} d_{A,x_i}^{-\alpha_A},
\end{eqnarray}
where $G^{\rm A}(d_{A,x_i})=G_{\rm m}^{\rm A}$ with a probability of $q_{\rm A}$ and $G^{\rm A}(d_{A,x_i})=g_{\rm s}^{\rm A}$ with a probability of $1-q_{\rm A}$.

 We provide the list of symbols in Table \ref{tab:notations}.
\begin{table}[!t]
	\caption{List of Symbols}
	\label{tab:notations}
		\fontsize{10}{10}\selectfont
	\centering
	\begin{tabular}{c |c }
		\hline \hline   \bfseries Notation &  \bfseries Description  \\  \hline
		$h_{\rm UT}$   & Height of the aerial user with reference to the terrestrial-BSs\\ \hline
		$h_{\rm UA}$   & Height of the aerial user with reference to the aerial-BSs \\ \hline
  	   $h_{\rm U}$   & Height of the aerial user with reference to the ground level \\ \hline	
  	   	$P_T,\ P_A$   & Transmit power of terrestrial and aerial BSs, respectively \\ \hline
	    $r_{\rm D}$   & Radius of the circle where aerial-BSs are distributed \\ \hline  	   
	    ${\rm H_C}$   & Height of the cylinder  where aerial-BSs are distributed in the 3-D setup \\ \hline 	    
		$\Phi_{\rm T},\ \Phi_{\rm A},\ \Phi_{\rm L},\ \Phi_{\rm N}$   & \makecellC{Tier of terrestrial-BSs, aerial-BSs, LoS terrestrial-BSs,\\ or NLoS terrestrial-BSs, respectively} \\ \hline
		$\lambda_{\rm T}$   & Density of terrestrial-BSs \\ \hline
		$P_{\rm L}(\cdot),\ P_{\rm N}(\cdot)$   & \makecellC{Probability of the aerial user being in LoS and NLoS\\with terrestrial-BSs, respectively} \\ \hline
		$G_{\rm m}^{\rm A},\ g_{\rm s}^{\rm A}$   & \makecellC{Gain of the aerial-BS's antenna mainlobe and sidelobe, respectively} \\ \hline
		$G_{\rm m}^{\rm T},\ g_{\rm s}^{\rm T}$   & \makecellC{Gain of the terrestrial-BS's antenna mainlobe and sidelobe, respectively} \\ \hline
		$q_{\rm A}$   & \makecellC{Probability that the mainlobe of the interfering aerial-BS \\ is directed toward the aerial user} \\ \hline
		$m_{L},\ m_{N},\ m_{A}$   & \makecellC{Nakagami-$m$ fading parameter for LoS terrestrial-BS,\\ NLoS terrestrial-BS, and aerial-BS, respectively} \\ \hline
			$\alpha_{L},\ \alpha_{N},\ \alpha_{A}$   & \makecellC{Path-loss exponent parameter for LoS terrestrial-BS,\\ NLoS terrestrial-BS, and aerial-BS, respectively} \\ \hline
	$d_{L,x_i},\ d_{N, x_i},\ d_{A,x_i}$   & \makecellC{Distance between the aerial user and an LoS terrestrial-BS,\\ NLoS terrestrial-BS, or aerial-BS, respectively, located at $x_i$} \\ \hline
		$R_{L},\ R_{N},\ R_{A}$   & \makecellC{Distance between the aerial user and its nearest LoS terrestrial-BS,\\ NLoS terrestrial-BS, or aerial-BS, respectively} \\ \hline
			$\tilde{R}_{L},\ \tilde{R}_{N},\ \tilde{R}_{A}$   & \makecellC{Distance between the aerial user and its serving BS \\assuming the that the aerial user is associated with an LoS terrestrial-BS,\\ NLoS terrestrial-BS, or aerial-BS, respectively} \\ \hline
        $\mathcal{E}_{\rm L},\ \mathcal{E}_{\rm N},\ \mathcal{E}_{\rm A}$   & \makecellC{Event that describes an aerial user's serving BS as LoS terrestrial-BS, \\ NLoS terrestrial-BS, or aerial-BS, respectively} \\ \hline			
		${\rm H}_{\rm L}^{x_i},\ {\rm H}_{\rm N}^{x_i},\ {\rm H}_{\rm A}^{x_i}$   & \makecellC{Channel power gain between the aerial user and  an LoS terrestrial-BS,\\ NLoS terrestrial-BS, or aerial-BS, respectively, located at $x_i$} \\ \hline			$I_{L},\ I_{N},\ I_{A}$   & \makecellC{Interference  at the aerial user from LoS terrestrial-BS,\\ NLoS terrestrial-BS, and aerial-BS, respectively} \\ \hline
	$\mathcal{A}_{\rm L},\ \mathcal{A}_{\rm N},\ \mathcal{A}_{\rm A}$   & \makecellC{Probability of association between aerial user and LoS terrestrial-BS,\\ NLoS terrestrial-BS, or aerial-BS, respectively} \\ 
	 \hline\hline
		
	\end{tabular}
\end{table}
\section{Relevant Distance  Distributions and Association Probabilities}
In this section, we begin by deriving  several relevant distance distributions that will be useful in deriving the association probabilities and  Laplace transforms of the interference powers given in (\ref{eq:int}). Afterwards, simplified expressions for the association probabilities and the terrestrial  interference's Laplace transforms are presented under the assumption that all  C2A links are in LoS condition.
\vspace{-0.4cm}
\subsection{Distance distributions of the nearest BSs}
In Lemma \ref{lem:1}, we present the distribution of the distances between the aerial user and its nearest aerial-BS, LoS terrestrial-BS, and NLoS terrestrial-BS.
\begin{lemma}
	\label{lem:1}
The PDF and the cumulative distribution function (CDF) of $R_\nu, \nu \in \{\rm L,N\}$ are given by
\vspace{-0.3cm}
\begin{eqnarray}
\label{eq:pdfRNRL}
\!\!\!\!\!\!\!\!\!\!\!\!\!\!\!\!\!\!f_{R_\nu}(r)&=&2 \pi \lambda_{\rm T} r P_\nu\!\left(\!\!\sqrt{\!r^2\!-\!h_{\rm UT}^2}\right)\text{exp}\!\left(\!-2 \pi \lambda_{\rm T}\!\!\int_{h_{\rm UT}}^{r}\!\!\!t P_\nu\!\left(\!\sqrt{\!t^2\!-\!h_{\rm UT}^2}\right)\!dt\!\right), r \geqslant h_{\rm UT}\\
\label{eq:cdfRNRL}
\!\!\!\!\!\!F_{R_\nu}(r)&\!=\!&1-\text{exp}\!\left(-2 \pi \lambda_{\rm T}\int_{h_{\rm UT}}^{r}tP_\nu\left(\sqrt{t^2-h_{\rm UT}^2}\right)dt\right), \hspace{55pt} r \geqslant h_{\rm UT}.
\end{eqnarray}
\end{lemma}

\textit{Proof:} See Appendix \ref{ap:RL}.

 The PDF and the CDF of $R_A$ are given by \cite[eq. (7)]{chetlur}
\begin{equation}
\label{eq:frA}
f_{R_A}(r)=N \left(\frac{2 r}{r_{\rm D}^2}\right)\left(\frac{d^2-r^2}{r_{\rm D}^2}\right)^{N-1},h_{\rm UA}\leqslant r \leqslant d \\
\end{equation}
\begin{equation}
\label{eq:FrA}
\!\!\!\!\!\!\!\!\!\!\!\!\!\!\!\!\!\!F_{R_A}(r)=\left\{
	\begin{array}{ll}
	0,& r\leqslant h_{\rm UA}\\
1- \left(\frac{d^2-r^2}{r_{\rm D}^2}\right)^{N}, & h_{\rm UA}\leqslant r \leqslant d\\
	1,& r\geqslant d,
	\end{array}
	\right.
\end{equation}
where $d=\sqrt{r_{\rm D}^2+h_{\rm UA}^2}$.
\vspace{-0.5cm}
\subsection{Distances of the nearest interfering BSs}
In Table \ref{my-label3}, we  summarize the distances $\tau_{\nu|\mathcal{E}_\nu}(r), \nu \in \{\rm L,N,A\}$ between the typical aerial user and the nearest interfering BSs  from the three tiers $\Phi_{\rm L},\Phi_{\rm N},\Phi_{\rm A}$ where the serving BS is at a distance of $r$. More specifically, the set of distances where $r \geqslant \tau_{\nu|\mathcal{E}_\nu}(r)$ quantify the region over which the interfering BSs from $\Phi_\nu$ exist. Note that $N'$ in Table \ref{my-label3} quantifies the number of interfering aerial-BSs.
\vspace{-0.5cm}

\begin{table*}[!t]
	\caption{Interferer distances}
	\label{my-label3}
	\centering
	\begin{tabular}{c |c| c| c| c}
		\hline\hline    Condition &  $\tau_{L|\mathcal{E}_{\rm \nu}}(r)$   &  $\tau_{N|\mathcal{E}_{\rm \nu}}(r)$ &  $\tau_{A|\mathcal{E}_{\rm \nu}}(r)$ &  N'  \\  \hline
		$\mathcal{E}_{\rm L}$ ($x_0 \in \Phi_{\rm L}$)   & $r$     & 
		
	 \bigcell{c}{$\left\{
	 	\begin{array}{ll}
	 	h_{\rm UT},& \!\!\!\!\!\!\!\!\!\!\!\!\!\!\!\!\!\!\!\! h_{\rm UT}\leqslant r\leqslant \zeta_{\rm N}^{\rm L}\\
	 	\left(\frac{\eta_{N}r^{\alpha_L}}{\eta_{L}}\right)^{\frac{1}{\alpha_{N}}},& r\geqslant\zeta_{\rm N}^{\rm L},
	 	\end{array}
	 	\right.$ \\ 	where $\zeta_{\rm N}^{\rm L}=\left(\frac{\eta_{L}h_{\rm UT}^{\alpha_{N}}}{\eta_{N}}\right)^{\frac{1}{\alpha_{L}}}$}& 
		 \bigcell{c}{$\left\{
		 	\begin{array}{ll}
		 	\!\!\!h_{\rm UA}, & \!\!\!\!\!\!\!\!\!\!\!\!\!\!\!\!\!\!\!\! h_{\rm UT}\leqslant \zeta_{\rm A}^{\rm L}(h_{\rm UA})\\
		 	\!\!\!\left(\frac{\mu_Ar^{\alpha_L}}{\mu_L}\right)^{\frac{1}{\alpha_{A}}}, \\
		 	& \!\!\!\!\!\!\!\!\!\!\!\!\!\!\!\!\!\!\!\!\zeta_{\rm A}^{\rm L}(h_{\rm UA})\leqslant\! r\!\leqslant \zeta_{\rm A}^{\rm L}(d),
		 	\end{array}
		 	\right.$  \\ 	where $\zeta_{\rm A}^{\rm L}(x)=\left(\frac{\mu_Lx^{\alpha_A}}{\mu_A}\right)^{\frac{1}{\alpha_{L}}}$}& N          \\ \hline
		$\mathcal{E}_{\rm N}$ ($x_0 \in \Phi_{\rm N}$ )  &   $\left(\frac{\mu_Lr^{\alpha_N}}{\mu_N}\right)^{\frac{1}{\alpha_{L}}}$    &   $r$  &   $h_{\rm UA}$ &   N  \\ \hline
		$\mathcal{E}_{\rm A}$ ($x_0 \in \Phi_{\rm A}$)   & 
		
		\bigcell{c}{$\left\{
			\begin{array}{ll}
			h_{\rm UT}, \!\!\!&\!\!\!\!\!\!\!\!\!\!\!\!\!\!\!\!\!\!\!\!\!\!\!\! h_{\rm UA}\leqslant r\leqslant \zeta_{\rm L}^{\rm A}\\
			\left(\frac{\mu_Lr^{\alpha_A}}{\mu_A}\right)^{\frac{1}{\alpha_{L}}}, \!\!\!\\&\!\!\!\!\!\!\!\!\!\!\!\!\!\!\!\!\!\!\!\!\!\!\!\!\zeta_{\rm L}^{\rm A}  \leqslant r\leqslant d.
			\end{array}
			\right.$ \\ 	where $\zeta_{\rm L}^{\rm A}=\left(\frac{\mu_Ah_{\rm UT}^{\alpha_L}}{\mu_L}\right)^{\frac{1}{\alpha_{A}}}$} &    $h_{\rm UT}$ &   $r$ &   N-1  \\ \hline\hline
		
	\end{tabular}
\end{table*}

\subsection{Association probabilities}
As stated previously, the typical aerial user is associated with an aerial-BS, LoS terrestrial-BS, or NLoS terrestrial-BS with probabilities given in the following lemmas.
\begin{lemma}
	\label{lem:AL}
The probability that the typical aerial user is associated with an LoS terrestrial-BS is given by
\vspace{-0.3cm}
\begin{equation}
\label{eq:AL}
\mathcal{A}_{\rm L}=\Xi^{\rm L}_{\rm N} \times \Xi^{\rm L}_{\rm A} ,
\end{equation}
where
\begin{equation}
\label{eq:Xi}
\Xi^{\rm L}_{\rm N}=\left\{
\begin{array}{ll}F_{R_L}(\zeta_{\rm N}^{\rm L})\!+\!\displaystyle\int_{\zeta_{\rm N}^{\rm L}}^{\infty}F_{R_N}^{(c)}\left(\!\!\left(\frac{\eta_{N} r^{\alpha_L}}{\eta_{L}}\right)^{\frac{1}{\alpha_{N}}}\!\!\right) f_{R_L}(r)dr,&  h_{\rm UT} \leqslant \zeta_{\rm N}^{\rm L}\\
\displaystyle\int_{\zeta_{\rm N}^{\rm L}}^{\infty}F_{R_N}^{(c)}\left(\left(\frac{\eta_{N} r^{\alpha_L}}{\eta_{L}}\right)^{\frac{1}{\alpha_{N}}}\right) f_{R_L}(r)dr,&h_{\rm UT} \geqslant \zeta_{\rm N}^{\rm L},
\end{array}
\right.
\end{equation}
and,
\begin{equation}
\label{eq:XiA}
\Xi^{\rm L}_{\rm A}=\left\{
\begin{array}{ll}
1+\displaystyle\int_{\zeta}^{\zeta_{\rm A}^{\rm L}(d)}\!\!\!\!F^{(c)}_{R_A}\!\left(\!\left(\!\frac{\mu_A r^{\alpha_L}}{\mu_L}\!\right)^{\frac{\!1}{\!\alpha_{A}}}\!\right)\!f_{R_L}\!(r)dr-F_{R_L}^{(c)}(\zeta),& h_{\rm UT}\!\leqslant\!\zeta_{\rm A}^{\rm L}(d)\\
0,& h_{\rm UT}\geqslant \zeta_{\rm A}^{\rm L}(d),
\end{array}
\right.
\end{equation}

with $\zeta=\text{arg}\max\Bigl\{h_{\rm UT},\zeta_{\rm A}^{\rm L}(h_{\rm UA})\Bigr\}$.
\end{lemma}

\textit{Proof:} See Appendix \ref{ap:Lemma1}.

\begin{lemma}
	The probability of the typical aerial user being associated with an aerial-BS is given by
	\vspace{-0.3cm}
	\begin{equation}
	\label{eq:AA}
	\mathcal{A}_{\rm A}=\Xi^ {\rm A}_{\rm N} \times \Xi^{\rm A}_{\rm L},
	\end{equation}
	where
	\begin{equation}
	\label{eq:XiAN}
	\Xi^ {\rm A}_{\rm N}\hspace{-3pt}=\hspace{-3pt}\left\{
	\begin{array}{ll}
\hspace{-5pt}\displaystyle\int_{h_{\rm UA}}^{d}\!F_{R_N}^{(c)}\left(\!\left(\!\frac{\mu_N r^{\alpha_A}}{\mu_A}\!\right)^{\frac{\!1}{\!\alpha_{N}}}\!\right)\!f_{R_A}\!(r)dr,
& \zeta_{\rm N}^{\rm A}\leqslant\! h_{\rm UA}\\
	F_{R_A}(\zeta_{\rm N}^{\rm A})\!+\!\!\displaystyle\int_{\zeta_{\rm N}^{\rm A}}^{d}\!\!\!F_{R_N}^{(c)}\left(\!\left(\!\frac{\mu_N r^{\alpha_A}}{\mu_A}\!\right)^{\frac{\!1}{\!\alpha_{N}}}\!\right)\!f_{R_A}\!(r)dr,
	& h_{\rm UA} \!\leqslant\! \zeta_{\rm N}^{\rm A}\!\leqslant \! d \\
	1, & \zeta_{\rm N}^{\rm A}\geqslant\! d, 
	\end{array}
	\right.
	\end{equation}
	where $\zeta_{\rm N}^{\rm A}=\left(\frac{\mu_A h_{\rm UT}^{\alpha_N}}{\mu_N}\right)^{\frac{1}{\alpha_{A}}}$, and, 
	\begin{equation}
	\label{eq:XiAL}
	\Xi^{\rm A}_{\rm L}\!=\!\left\{
	\begin{array}{ll}
	\displaystyle\int_{h_{\rm UA}}^{d}\!F_{R_L}^{(c)}\left(\!\left(\!\frac{\mu_L r^{\alpha_A}}{\mu_A}\!\right)^{\frac{\!1}{\!\alpha_{L}}}\!\right)\!f_{R_A}\!(r)dr,
&  \zeta_{\rm L}^{\rm A}\leqslant\! h_{\rm UA}\\
	F_{R_A}(\zeta_{\rm L}^{\rm A})\!+\!\!\displaystyle\int_{\zeta_{\rm L}^{\rm A}}^{d}\!\!F_{R_L}^{(c)}\left(\!\left(\!\frac{\mu_L r^{\alpha_A}}{\mu_A}\!\right)^{\frac{\!1}{\!\alpha_{L}}}\!\right)\!f_{R_A}\!(r)dr,
& h_{\rm UA} \leqslant \!\zeta_{\rm L}^{\rm A}\!\leqslant\! \! d \\
	1,
	& \zeta_{\rm L}^{\rm A}\geqslant\! d. \\
	\end{array}
	\right.
	\end{equation}
\end{lemma}
\textit{Proof:}   The proof follows the same steps as that of Lemma \ref{lem:AL}; therefore, it is omitted here.

 Finally, the probability of the aerial user being associated with an NLoS terrestrial-BS is given by $ \mathcal{A}_{\rm N}=1-\mathcal{A}_{\rm L}- \mathcal{A}_{\rm A}$.
 
 \subsection{Conditional distance distributions of the serving BS}
 In this section, we present the distribution of the distances between the typical aerial user and its serving BS, where the latter is an  LoS terrestrial-BS, NLoS terrestrial-BS, or aerial-BS.
 \begin{lemma}\label{lma:DisDistanceL}
 	Where the aerial user is associated with an LoS terrestrial-BS (i.e., the event $\mathcal{E}_{\rm L}$ occurs), the distribution of the distance $\tilde{R}_L$ between the aerial user and the serving LoS terrestrial-BS  is given by
 	\begin{equation}
 	\label{eq:rltild}
	\!\!\!f_{\tilde{R}_L|\mathcal{E}_{\rm L}}(r)\!\!=\!\!\left\{\!\!\!
	\begin{array}{ll}
	\frac{1}{\mathcal{A}_{\rm L}}F^{(c)}_{R_N}\left(\tau_{N|\mathcal{E}_{\rm L}}(r)\right)\times F^{(c)}_{R_A}\left(\tau_{A|\mathcal{E}_{\rm L}}(r)\right)\!\!\times\!\!f_{R_L}(\!r\!),
	& r \in \mathfrak{R}_{\rm L}\\
0, & \text{otherwise},
	\end{array}
	\right.
\end{equation}
 	where $\mathfrak{R}_{\rm L}=[h_{\rm UT},\zeta_{\rm A}^{\rm L}(d)]$. Note that according to (\ref{eq:XiA}), $\mathcal{A}_{\rm L}=0$ outside $\mathfrak{R}_{\rm L}$.
 \end{lemma}
 
 \textit{Proof:}   See Appendix \ref{ap:Lemma3}.

 \begin{lemma}
 	Where the aerial user is associated with an aerial-BS (i.e., the event $\mathcal{E}_{\rm A}$ occurs), the distribution of the distance $\tilde{R}_A$ between the aerial user and the serving aerial-BS is given by
 	 	\begin{equation}
 	\label{eq:ratild}
	\!\!\!f_{\tilde{R}_A|\mathcal{E}_{\rm A}}(r)\!\!=\!\!\left\{\!\!\!
	\begin{array}{ll}
\frac{1}{ \mathcal{A}_{\rm A}}F^{(c)}_{R_N}\left(h_{\rm UT}\right)\times F^{(c)}_{R_L}\left(\tau_{L|\mathcal{E}_{\rm A}}(r)\right)\times f_{R_A}(r),
	& r \in \mathfrak{R}_{\rm A}\\
0, &\text{otherwise},
	\end{array}
	\right.
\end{equation}
where $\mathfrak{R}_{\rm A}=[h_{\rm UA}, d]$.
 \end{lemma}
 \textit{Proof:} The proof follows the same steps as that of Lemma \ref{lma:DisDistanceL}; therefore, it is omitted here.

 \begin{lemma}
 	Where the aerial user is associated with an NLoS terrestrial-BS (i.e., the event $\mathcal{E}_{\rm N}$ occurs),
 	the distribution of the distance $\tilde{R}_N$ between the aerial user and the serving NLoS terrestrial-BS is given by
\begin{equation}
 	\label{eq:rntild}
	f_{\tilde{R}_N|\mathcal{E}_{\rm N}}(r)=\left\{\!\!\!
	\begin{array}{ll}
\frac{1}{ \mathcal{A}_{\rm N}}F^{(c)}_{R_L}\left(\tau_{L|\mathcal{E}_{\rm N}}(r)\right)\times F^{(c)}_{R_A}\left(h_{\rm UA}\right)\times f_{R_N}(r),
	& r \in \mathfrak{R}_{\rm N}\\
0, &\text{otherwise},
	\end{array}
	\right.
\end{equation}
where $ \mathfrak{R}_{\rm N}=[h_{\rm UT}, d]$.
 \end{lemma}
 \vspace{-0.7cm}
\subsection{Laplace transform of the aggregated interference}
Depending on the spectrum sharing policy, the aerial user may receive interference signals from aerial-BSs and terrestrial-BSs whose Laplace transforms are characterized in the following lemmas. 
\vspace{-0.3cm}
\begin{lemma}
	\label{lem:LTT}
	Conditioned on the event  $\mathcal{E}_{\nu}$, $\nu \in \{\rm L,N, A\}$, the Laplace transform of the terrestrial interference power, $I_L+I_N$, is given by
	\vspace{-0.2cm}
	\begin{equation}
	\label{eq:Li}
	\mathcal{L}_{(I_N+I_L)|\mathcal{E}_{ \nu}}(s  )\!=\!
	\!\!\!\!\!\prod_{\omega\in\{\rm L,N\}}\!\!\!\!\!\exp\Big(\!\!-\!2\pi \lambda_{\rm T}\int_{\tau_{\omega|\mathcal{E}_{\rm \nu}}(r )}^{\infty}\!\!\Big(\!1\!-\!\left(\!\frac{\!\!\!m_\omega}{m_\omega\!+\!s\mu_\omega t^{-\alpha_{\omega}}}\right)^{\!\!\!m_\omega}\!\Big)tP_{\rm \omega}\!\!\left(\!\!\sqrt{\!t^2\!-\!h_{\rm UT}^2}\right)\!dt\!\Big).
	\end{equation}
	
\end{lemma}
\textit{Proof:} See Appendix \ref{ap:LI}.
\begin{lemma}
	\label{lem:LTA}
	Conditioned on the event  $\mathcal{E}_{\nu}$, $\nu \in \{\rm L,N, A\}$, the Laplace transform of the aerial interference power, $I_A$, is given by
		\begin{eqnarray}
	\label{eq:Li1}
	\mathcal{L}_{I_{ A}|\mathcal{E}_{\rm \nu}}(s )&=&\sum_{i=0}^{N'}\binom{N'}{i}\!\left(\!\frac{2 m_A^{m_A}(sP_A\eta_A)^{-m_A}}{\alpha_{A}\Gamma(m_A)(d^2-r^2)}\right)^{N'}\times  \left[q_{\rm A}\left(\Omega(d,G_{\rm m}^{\rm A})-\Omega(\tau_{A|\mathcal{E}_{\rm \nu}}(r)),G_{\rm m}^{\rm A})\right)\right]^{N'-i}\nonumber\\
&&\times\left[(1-q_{\rm A})\left(\Omega(d,g_{\rm s}^{\rm A})-\Omega(\tau_{A|\mathcal{E}_{\rm \nu}}(r),g_{\rm s}^{\rm A})\right)\right]^{i},
	\end{eqnarray}
	where $\Omega[\cdot,\cdot]$ is given below by
	\begin{equation}
\label{eq:Omeg}
\Omega(x,g)=\frac{x^{\alpha_A m_A+2}}{g^{m_A}}{\rm G}_{2,2}^{1,2}\Biggl[\!\frac{m_Ax^{\alpha_{A}}}{s P_A g \eta_{A}}\! \Bigg\vert\! \ \ {\!1\!-\!m_A\!-\!\frac{2}{\alpha_A},1\!-\!m_A\! \atop 0,-m_A-\frac{2}{\alpha_A}}\Biggr],
	\end{equation}
	where ${\rm G}[\cdot]$ is the Meijer-G function given in \cite[eq (9.301)]{grad}.
\end{lemma}

\textit{Proof:} See Appendix \ref{ap:LIA}.
\subsection{Simplified C2A channel model}
\label{sub:LOS}
The previous results (association probabilities and  Laplace transform of interference) require numerical evaluations of multiple integrals. These expressions can be simplified by noting that the aerial user may hover at elevated heights resulting in LoS transmissions with the terrestrial-BSs. Therefore, we assume that the C2A links are in LoS condition with the aerial user. The validity of this assumption will be investigated with simulations in Section \ref{sec:SimuResu}. Under this assumption, the PDF and CDF of the distance between the aerial user and the nearest LoS terrestrial-BS are given by \cite[eqs. (8) and (9)]{chetlur2019coverage}
\vspace{-0.3cm}
\begin{eqnarray}
\label{eq:RT}
f_{R_L}(r)&=&2 \pi \lambda_{\rm T} r \exp\left(-\pi \lambda_{\rm T} \left(r^2-h_{\rm UT}^2\right)\right), r \geqslant h_{\rm UT}\\
F_{R_L}(r)&=&1-\exp\left(-\pi \lambda_{\rm T} \left(r^2-h_{\rm UT}^2\right)\right), \hspace{11pt}\hspace{3pt}r \geqslant h_{\rm UT}.
\end{eqnarray}

\begin{corollary}
\label{cor:1}
With the assumption that the aerial user is in LoS with the terrestrial-BSs, the association probability with a terrestrial-BS given in (\ref{eq:AL}) can be written as in (\ref{eq:AT}) at the top of the next page, where $\Upsilon(x)=\Gamma\left[\frac{\alpha_{L}}{\alpha_A}i +1,\pi \lambda_{\rm T} x^2 \right]$, with $\Gamma[\cdot,\cdot]$ denoting the upper incomplete Gamma function \cite[eq (8.358.2)]{grad}. Thus, the probability of association with an aerial-BS is given by $\mathcal{A}_{\rm A}=1-\mathcal{A}_{\rm L}$.

\end{corollary}

\textit{Proof:} Substitute (\ref{eq:RT}) into (\ref{eq:XiA}) and apply the Binomial theorem. Then, using the definition of the upper incomplete Gamma function and  some manipulations, we obtain (\ref{eq:AT}).

\begin{corollary}
	\label{cor:2}
	The Laplace transform of the terrestrial interference, $\mathcal{L}_{I_L|\mathcal{E}_{ \nu}}(s)$, where $\nu \in \{\rm L,A\}$, is given by
\begin{figure*}[!t]
	\begin{equation}
	\label{eq:AT}
	\mathcal{A}_{\rm L}\!=\!\!\left\{
	\begin{array}{ll}
	1\!\!-\exp\left(-\pi \lambda_{\rm T} \left(\zeta^2-h_{\rm UT}^2\right)\right)+\frac{\pi \lambda_{\rm T} h_{\rm UT}^2}{r_{\rm D}^{2 N}}\displaystyle\sum_{i=0}^{N}\binom{N}{i}(-1)^{i} d^{2(N-i)}\left(\frac{\mu_A}{\mu_L}\right)^{\frac{2 i}{\alpha_A}} \left(\pi \lambda_{\rm T}\right)^{-\frac{\alpha_{L} i}{\alpha_{A}}}\\
	\times\left[\Upsilon(\zeta)-\Upsilon(\zeta_{\rm A}^{\rm L}(d))\right], &\!\!\!\!\!\!\!\!\!\!\!\!\!\!\!\!\!\!\!\!\!\!\!\!\!\!\!\!\!\!\!\!\!\!\!\!\!\!\!\!\!\!h_{\rm UT}\!\leqslant\!\zeta_{\rm A}^{\rm L}(d)\\
	0,&\!\!\!\!\!\!\!\!\!\!\!\!\!\!\!\!\!\!\!\!\!\!\!\!\!\!\!\!\!\!\!\!\!\!\!\!\!\!\!\!\!\!h_{\rm UT}\geqslant \zeta_{\rm A}^{\rm L}(d).
	\end{array}
	\right.
	\end{equation}
	\hrule
\end{figure*}	
\vspace{-0.3cm}
		\begin{equation}
	\label{eq:LT}
\!\!\!\!\!\!\mathcal{L}_{I_L|\mathcal{E}_{ \nu}}(s)=\exp\Big(\!\!-\!2\pi \lambda_{\rm T}\sum_{i=0}^{m_L-1}\left(\frac{s \mu_L}{m_L}\right)^{1+i-m_L}\!\!\!\!\!\!\frac{1}{\alpha_{L}\Gamma[m_L-i]}\tau_{L|\mathcal{E}_{\rm \nu}}(r)^{\alpha_{L}(m_L-i-1)}\Theta(s,\tau_{L|\mathcal{E}_{\rm \nu}}(r))\Big),
	\end{equation}
	where $\Theta(\cdot,\cdot)$ is defined as
	\begin{equation}
	\label{eq:thet}
\Theta(s,r)={\rm G}_{2,2}^{2,1}\Biggl[\!\frac{m_Lr^{\alpha_{L}}}{s \mu_L}\! \Bigg\vert\! \ \! {1\!+\!i\!-\!m_L,2\!+\!i\!-\!m_L\!-\!\frac{2}{\alpha_L}\atop 1+i-m_L-\frac{2}{\alpha_L},0}\Biggr].
	\end{equation}
\end{corollary}
\textit{Proof:} Starting from (\ref{eq:Li}), substituting $P_L(t)=1,\quad t\geqslant h_{\rm UT}$  and applying the identity $a^n-b^n=(a-b)\sum_{i=0}^{n-1}a^i b^{n-1-i}$, \cite[eq (1.43)]{mathai}, and \cite[eq (7.811.2)]{grad} yield (\ref{eq:LT}).

\section{Coverage and Rate Analysis}

\subsection{Downlink coverage probability}
The coverage probability is defined as the probability of the \textsf{SINR} at the typical aerial user exceeding a predetermined threshold $T$. By applying the law of total probability, the coverage probability at the typical aerial user is given by
\begin{equation}
\label{eq:cov}
\mathcal{C}= \sum_{\nu \in\{\rm L,N,A\}}^{}\mathcal{A}_{\rm \nu} \times \mathcal{C}_{\rm \nu},
\end{equation}
where $\mathcal{A}_{\nu}$ and $\mathcal{C}_\nu,\nu \in\{\rm L,N,A\}$ denote the association probability with the $\Phi_\nu$ tier and the coverage probability given that the aerial user is associated with the BS from  $\Phi_\nu$, respectively.
\begin{theorem}
The coverage probability \label{th:cov} $\mathcal{C}_\nu$ conditioned on the events $\mathcal{E}_\nu, \nu \in\{\rm L,N,A\}$ is given by
\begin{equation}
\label{eq:Pnu}
\mathcal{C}_\nu=\sum_{k=0}^{m_\nu-1}\!\!\frac{(-1)^k}{k!}\!\!\left(\frac{m_\nu T}{\mu_\nu}\right)^k\!\!\displaystyle \!\!\!\int_{r\in \mathfrak{R}_\nu}^{}\!\!\!r^{k\alpha_{\nu}}\left[{\frac{\partial^k}{\partial s^k}}\mathcal{L}_{V|\mathcal{E}_{ \nu}}(s )\right]_{s=\frac{m_\nu T r^{\alpha_\nu}}{\mu_\nu}}f_{\tilde{R}_\nu|\mathcal{E}_\nu}(r)dr,
\end{equation}
where $V=I+\sigma^2$ and  $\mathcal{L}_{V|\mathcal{E}_{ \nu}}(s )$ is given by $\mathcal{L}_{V|\mathcal{E}_{ \nu}}(s )=\exp(-\sigma^2 s)\mathcal{L}_{I|\mathcal{E}_{ \nu}}(s )$.

\end{theorem}
 
 Note that under the N-OSS policy, $\mathcal{L}_{I|\mathcal{E}_{ \nu}}(s )=\mathcal{L}_{(I_N+I_L)|\mathcal{E}_{ \nu}}(s  )\times \mathcal{L}_{I_{A|\mathcal{E}_{ \nu}}}(s  )$. 
 
\textit{Proof:} See Appendix \ref{ap:cov}.
\subsection{Downlink average achievable rate}
In this section, we derive the average achievable rate of the typical aerial user. Following the same analysis as the coverage probability, the average ergodic rate is given by
\begin{equation}
\label{eq:rate}
\mathcal{R}=\sum_{\nu \in\{\rm L,N,A\}}^{}\mathcal{A}_{\rm \nu} \times \mathcal{R}_{\rm \nu},
\end{equation}
where $\mathcal{R}_\nu$ is the average achievable rate of the aerial user when it is associated with a BS from the tier $\Phi_{\nu}$.
\begin{theorem}
\label{th:rate}
The average achievable rate of the typical aerial user when it is associated with $\Phi_\nu$ tier is given by
\begin{equation}
\label{eq:Rnu}
\mathcal{R}_\nu=\sum_{k=0}^{m_\nu-1}\!\!\frac{(-1)^k}{\ln(2)k!}\!\!\left(\frac{m_\nu }{\mu_\nu}\right)^k\!\!\displaystyle \!\!\!\int_{0}^{\infty}\int_{r\in \mathfrak{R}_\nu}^{}\!\!\!r^{k\alpha_{\nu}}(e^t-1)^k\left[{\frac{\partial^k}{\partial s^k}}\mathcal{L}_{V|\mathcal{E}_{ \nu}}(s )\right]_{s=\frac{m_\nu (e^t-1) r^{\alpha_\nu}}{\mu_\nu}}f_{\tilde{R}_\nu|\mathcal{E}_\nu}(r)dr dt,
\end{equation}

\end{theorem}
\textit{Proof:} Using Shannon's theorem, the average achievable rate can be expressed as
\begin{eqnarray}
\label{eq:pr}
\mathcal{R}_\nu&=&\frac{1}{\ln(2)}\mathbb{E}_{\tilde{R}_\nu}\left[\mathbb{E}_\gamma\left[\ln(1+\gamma)\right]\right]\overset{(a)}{=}\frac{1}{\ln(2)}\mathbb{E}_{\tilde{R}_\nu}\left[\displaystyle \int_{0}^{\infty}\mathbb{P}\left(\gamma\geqslant e^t-1 \right)dt\right],\!\!\!\!\!\!
\end{eqnarray}
where (a) follows from  $\mathbb{E}[Z]=\int_{0}^{\infty}\mathbb{P}(Z>t)dt$. Substituting (\ref{eq:sinr}) into (\ref{eq:pr}) and following the same steps as those of Theorem  \ref{th:cov} yields (\ref{eq:Rnu}).
\subsection{Downlink coverage and average achievable rate approximations}
The evaluation of the results presented in (\ref{eq:Pnu}) and (\ref{eq:Rnu}) for the conditional coverage probabilities and average achievable rates involves higher order derivative of the Laplace transform, which leads to a more complex computation especially for large fading parameters, i.e., $m_\nu$, $\nu \in \{\rm L,N,A\}$. Thus, a lower bound of the Gamma distribution's CCDF is used instead of the exact evaluation. 

\begin{theorem}
\label{th:thappC}
The approximation of the conditional coverage probability  $\mathcal{C}_\nu^{\rm approx}$, $\nu \in\{\rm L,N,A\}$ is given by
\vspace{-0.3cm}
\begin{equation}
\label{eq:Pnuapp}
\mathcal{C}_\nu^{\rm approx}\approx	\sum_{k=1}^{m_\nu}\!\!{(-1)^{k+1}}\left(m_\nu \atop k \right)\!\displaystyle \!\!\int_{r\in \mathfrak{R}_\nu}^{}\mathcal{L}_{V|\mathcal{E}_{ \nu}}\left(\frac{k \rho_\nu T r^{\alpha_\nu}}{\mu_\nu}\right)
f_{\tilde{R}_\nu|\mathcal{E}_\nu}(r)dr,
\end{equation}
where  $\rho_\nu = m_\nu (m_\nu !)^{\frac{-1}{m_\nu}}$, with $m_\nu>1$.

\end{theorem}
\textit{Proof:} Using \cite{alzer1997}, the CCDF of a  Gamma-distributed random variable, $\rm H_\nu^{x_0}$,  can be tightly lower bounded by 
\vspace{-0.3cm}
\begin{equation}
\label{eq:GammApprox}
\mathbb{P}\left({\rm H}_{\nu}^{x_0} \geqslant \beta\right) \geqslant 1-[1-e^{-\rho_\nu \beta}]^{m_\nu}= \sum_{k=1}^{m_\nu}\!\!{(-1)^{k+1}}\left(m_\nu \atop k \right) e^{-k \rho_\nu \beta},
\end{equation}
where the equality holds when $m_\nu=1$.  Thus, substituting this upper bound to (c) in (\ref{eq:Pu})  along the same mathematical manipulations concludes the proof.

\begin{theorem}
The conditional average achievable rate  approximation  $\mathcal{R}_\nu^{\rm approx}$, $\nu \in\{\rm L,N,A\}$ is given by
\vspace{-0.3cm}
\begin{equation}
\label{eq:Rnuapp}
\mathcal{R}_\nu^{\rm approx}\approx	\frac{1}{\ln(2)}\sum_{k=1}^{m_\nu}\!\!{(-1)^{k+1}}\left(m_\nu \atop k \right)\displaystyle \!\!\!\int_{0}^{\infty}\!\displaystyle \!\!\int_{r\in \mathfrak{R}_\nu}^{}\mathcal{L}_{V|\mathcal{E}_{ \nu}}\left(\frac{k \rho_\nu  r^{\alpha_\nu} (e^{-t}-1)}{\mu_\nu}\right)
f_{\tilde{R}_\nu|\mathcal{E}_\nu}(r)dr dt.
\end{equation}
\end{theorem}
\textit{Proof:} The proof follows the same steps as that of Theorem \ref{th:thappC}; therefore, it is omitted here.
\vspace{-0.5cm}
\section{Results and Discussion}\label{sec:SimuResu}

\begin{table}[!t]
	\caption{Parameter values}
	\label{tab:param}
	\centering
	\begin{tabular}{c |c ||c| c||c| c}
		\hline \hline   \bfseries Parameter &  \bfseries Value   &  \bfseries Parameter &  \bfseries Value  &  \bfseries Parameter &  \bfseries value \\  \hline
		$(P_T,\ P_A)$   & (43, 30) dBm     & ($m_L,\ m_N,\ m_A$) & (2, 1, 2)      & N& 10   \\ \hline
		$r_{\rm D}$   & 2000 meters   & $(G_{\rm m}^{\rm T},\ g_{\rm s}^{\rm T})$   & (0, -15) dB     & ($\eta_{L},\ \eta_{N},\ \eta_{A}$)& (-3, -20, -1) dB  \\ \hline
		$(G_{\rm m}^{\rm A},\ g_{\rm s}^{\rm A})$   & (0, -20) dB     & C2A& LoS/NLoS & $\sigma^2$& -113 dBm\\ \hline
		$(\alpha_{L},\ \alpha_{N},\ \alpha_{A})$   & (2.5, 3.5, 2)  &$\lambda_{\rm T}$   &5     & $T$& 5 dB \\ \hline
	
		Policy   &N-OSS    & Environment& Urban & $\theta_{\rm B}^{\rm A}$   & $18^{\circ}$ \\ \hline 
		($h_{\rm U},\ h_{\rm A}$)& (50, 300) meters & $q_{\rm A}$   &0.1    & $h_{\rm T}$& 20 meters\\ \hline\hline
		
	\end{tabular}
\end{table}

In this section, we use Monte Carlo simulations to validate the analytical expressions. We also investigate the impact of several design parameters on the VHetNet's performance. Table \ref{tab:param} summarizes the parameters used in the simulations, unless otherwise indicated. It should be mentioned that we used parameters set by the 3GPP report in \cite{3gpp777} which limit the maximum hovering height for aerial users  to 300 meters. The height of an aerial-BS, however, can extend up to a few kilometers \cite{chetlur, alhouranilet,alzenadcoverage}.

\begin{figure}
\captionsetup{width=0.47\textwidth}
\centering
\begin{minipage}[t]{.5\textwidth}
  \centering
  \includegraphics[width=0.95\linewidth]{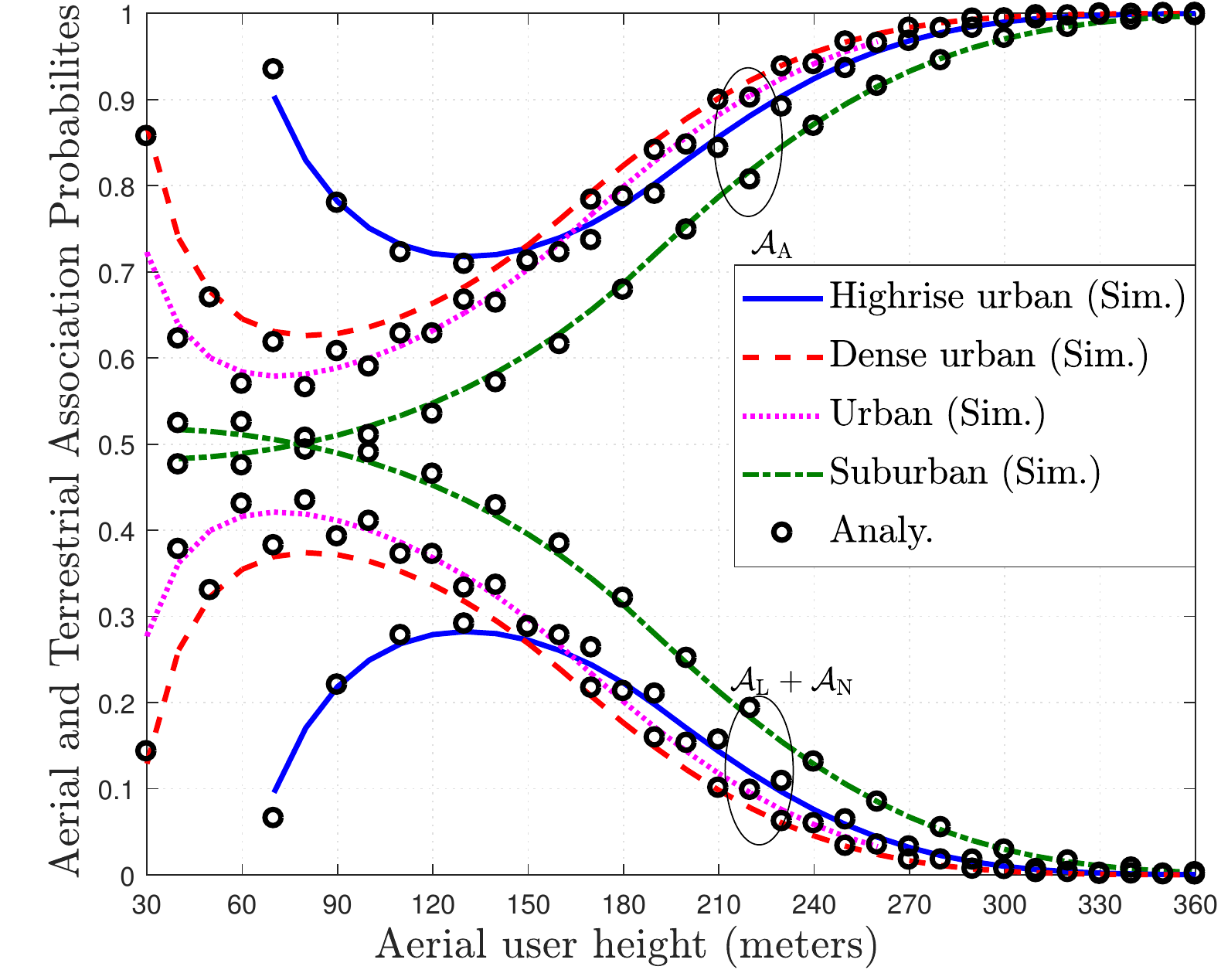}
  \captionof{figure}{Association probability versus aerial user height with $ h_{\rm A}=500 $ meters.}
  \label{ass}
\end{minipage}\hfill
\begin{minipage}[t]{.5\textwidth}
  \centering
  \includegraphics[width=1\linewidth]{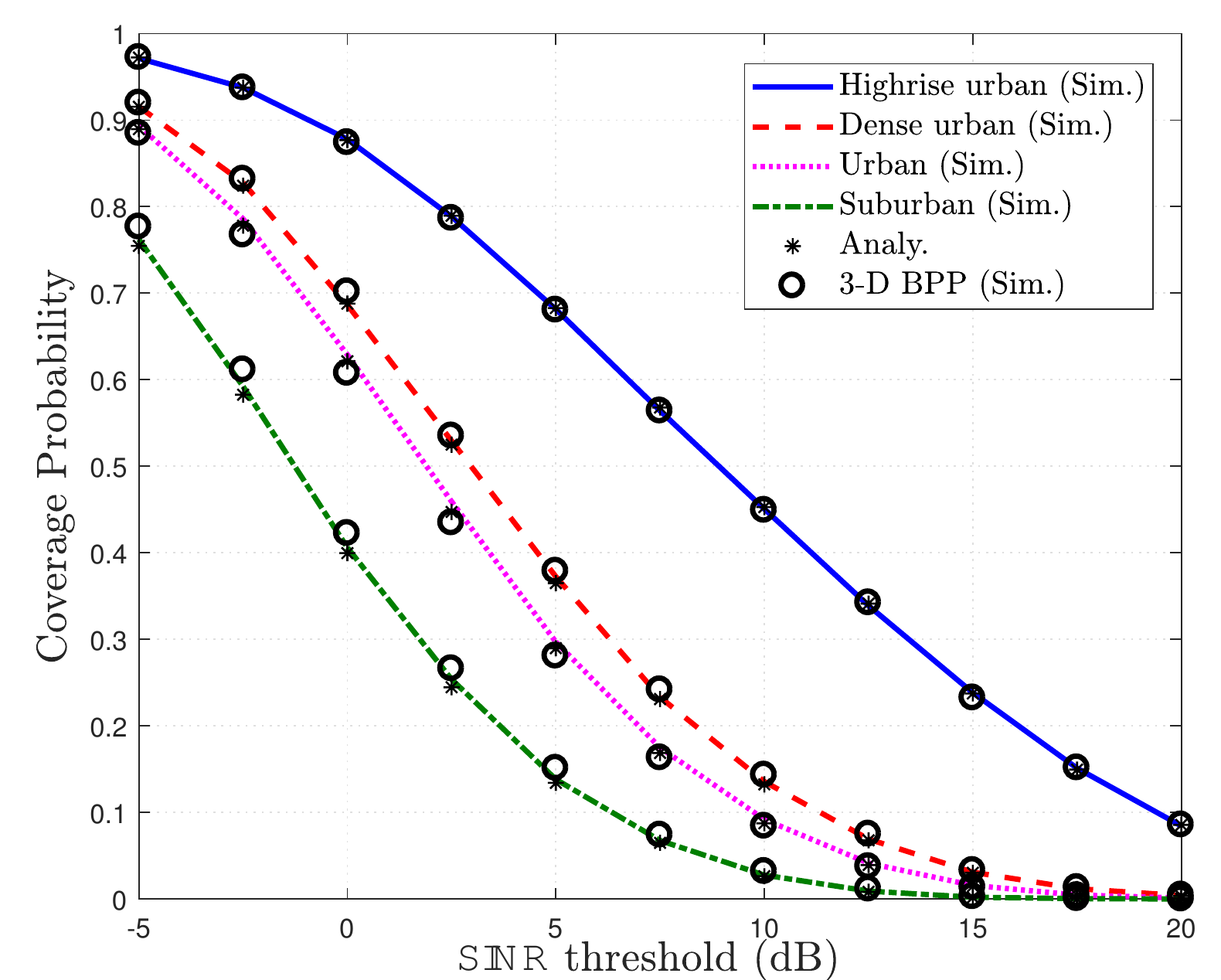}
  \captionof{figure}{Coverage probability versus \textsf{SINR} threshold: $h_{\rm U}=70$ meters and $ h_{\rm A}=200$ meters. For 3-D BPP scenario: $h_{\rm max}=225$ meters and $h_{\rm min}=175$ meters.}
  \label{th}
\end{minipage}
\end{figure}

In Fig. \ref{ass}, we compare aerial-BS and terrestrial-BS association probabilities with different aerial user heights and environments. Notably, at any aerial user height, $\mathcal{A}_{\rm A}+(\mathcal{A}_{\rm L}+\mathcal{A}_{\rm N})=1$. At low aerial user heights, the aerial user tends to be associated with an aerial-BS because the C2A links are dominated by NLoS links, which have poor channel conditions in comparison to  LoS A2A links. As the height increases, the terrestrial-BS association probability increases since more terrestrial-BSs come within LoS condition of the aerial user. However, with a further increase in height, the terrestrial-BS association probability starts to decrease because of the significant path-loss caused by the increasing distances between the aerial user and the terrestrial-BSs. Essentially, for a given environment and height, the aerial user probability of association to either a terrestrial or aerial network indicates the likelihood of the serving BS belonging to either of the two networks, respectively.

Fig. \ref{th} shows the coverage probability versus the \textsf{SINR} threshold for different environments. For a particular \textsf{SINR} threshold, as the environment becomes less dense, more terrestrial-BSs come into LoS with the aerial user (i.e., more interfering LoS terrestrial-BSs) which increases the interference. As discussed in Section II, the analytical tractability of a VHetNet with aerial-BSs at varying heights is quite challenging. Interestingly, aerial-BSs hovering at the same height $h_{\rm A}$ guarantee a comparable performance to VHetNets involving aerial-BSs with heights uniformly distributed between $h_{\rm max}$ and $h_{\rm min}$ with an average height of $h_{\rm A}$. The results presented in Fig. \ref{th} show that a small variation in the height of aerial-BSs does not affect the overall performance. This is because the aerial-BSs are distributed over the area of a disc with a radius that is much larger than the aerial-BS heights, which means that small variations in aerial-BS heights do not have a significant impact on the distribution of the distances separating the aerial user from the aerial-BSs. 

\begin{figure}
\captionsetup{width=0.47\textwidth}
\centering
\begin{minipage}[t]{.5\textwidth}
  \centering
  \includegraphics[width=1\linewidth]{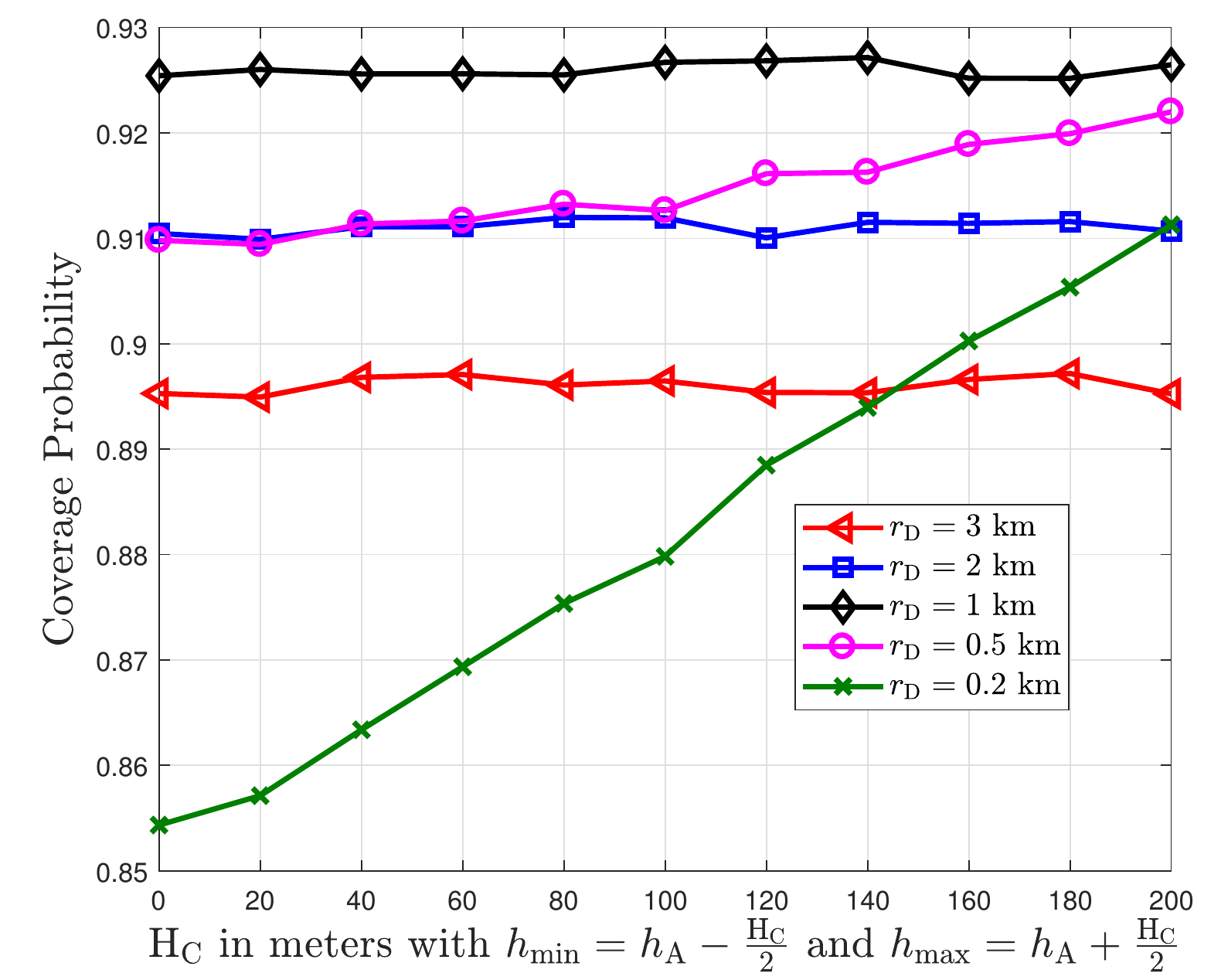}
  \captionof{figure}{Coverage probability for 3-D BPP scenario: $N$ aerial-BSs are uniformly distributed in a finite cylinder with height $\rm H_C$ and  radius $r_{\rm D}$ ($h_{\rm A}= 300$ meters and $T=-5$ dB).}
  \label{cyl}
\end{minipage}\hfill
\begin{minipage}[t]{.5\textwidth}
  \centering
  \includegraphics[width=1\linewidth]{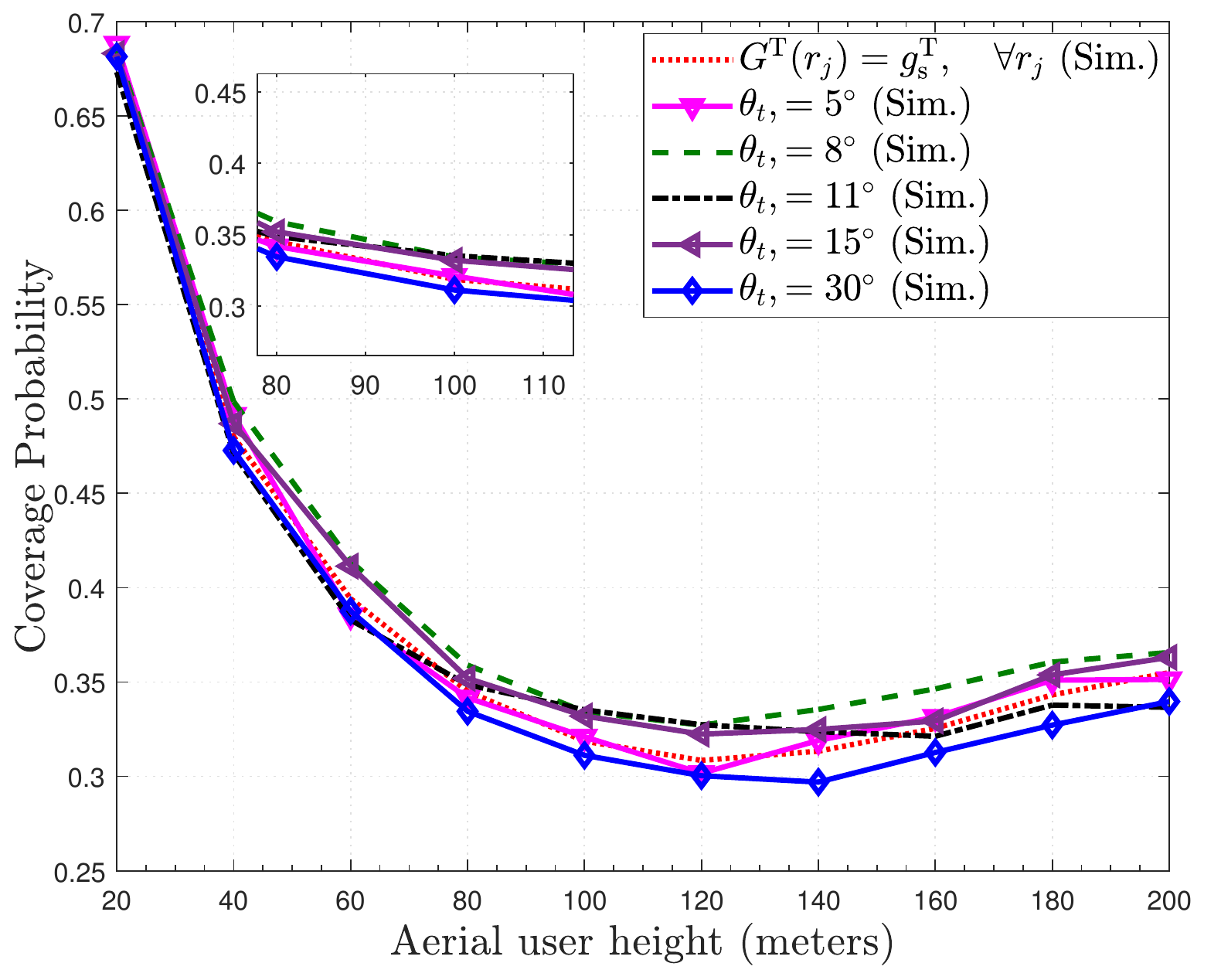}
  \captionof{figure}{Coverage probability versus aerial user height for different terrestrial-BS tilt-angles for $\theta_{\rm B}^{\rm T}=30^{\circ}$.}
  \label{cov_tilt}
\end{minipage}
\end{figure}

Fig. \ref{cyl} shows the performance of VHetNets where  aerial-BSs hover at different heights but are confined between two values: $h_{\rm min}=h_{\rm A}-\frac{\rm H_C}{2}$ and $h_{\rm max}=h_{\rm A}+\frac{\rm H_C}{2}$ as depicted in Fig. \ref{fig:sys3D}, where aerial-BSs are uniformly distributed within a finite cylinder of height $\rm H_C$ and radius $r_{\rm D}$. Fig. \ref{cyl} shows that the coverage probability using a 3-D BPP aerial network is the same as its counterpart assuming a 2-D BPP at a fixed height $h_{\rm A}$ (in this case $\rm H_C=0$) when $\rm H_C$ is much larger  than the aerial-BSs height, $h_{\rm A}$. However, the 2-D BPP assumption does not guarantee a likewise coverage probability under 3-D BPP when the radius over which the aerial-BSs are distributed uniformly, $r_{\rm D}$, is comparable to their heights, $h_{\rm A}$. Indeed, works in \cite{chetlur} and \cite{merwaday2016} indicate that aerial-BSs are distributed over a large area with a radius in the order of a few kilometers, whereas aerial-BS heights are in the order of a few hundred meters. Given this, we wish to emphasize that the framework presented in this work is applicable under the assumption that the radius of the disk $r_{\rm D}$ over which the aerial-BSs can be distributed is much larger than the cylinder height $\rm H_{\rm C}$. That is, an aerial network with a dense deployment of aerial-BSs, where their heights and spacing are comparable, does not necessarily exhibit the same performance as the aerial network in this work. This is because the assumption concerning order of magnitude differences between $r_{\rm D}$ and $\rm H_C$ is violated, and, therefore, revisiting the performance analysis using more comprehensive models is required.  Indeed, under the assumption $r_{\rm D} \gg \rm H_C$, Fig. \ref{cyl} shows that the complexity from assuming a 3-D BPP network can be avoided by adopting a 2-D BPP network with a fixed height without compromising the performance accuracy or correctness. Finally, we can affirm that the assumption of fixed heights of aerial-BSs is legitimate enough to enable the tractability of the analysis.

Fig. \ref{cov_tilt} illustrates the coverage probability for different terrestrial-BS tilt-angles. The performance of VHetNets that assume different $\theta_t$ is close enough to the performance of the system,  assuming that the aerial user is always serviced by the terrestrial-BS antenna sidelobes $G^{\rm T}(r_j)=g_{\rm s}^{\rm T}$,  $\forall\ r_j$. This observation illustrates well the assumption that  terrestrial-BSs cover  aerial users mainly with their antenna sidelobes, since these are usually down-tilted toward terrestrial users. Further, as we pointed out, this assumption is the key enabler of the tractability  of the association probabilities analysis.

\begin{figure}
\captionsetup{width=0.47\textwidth}
\centering
\begin{minipage}[t]{.5\textwidth}
  \centering
  \includegraphics[width=0.95\linewidth]{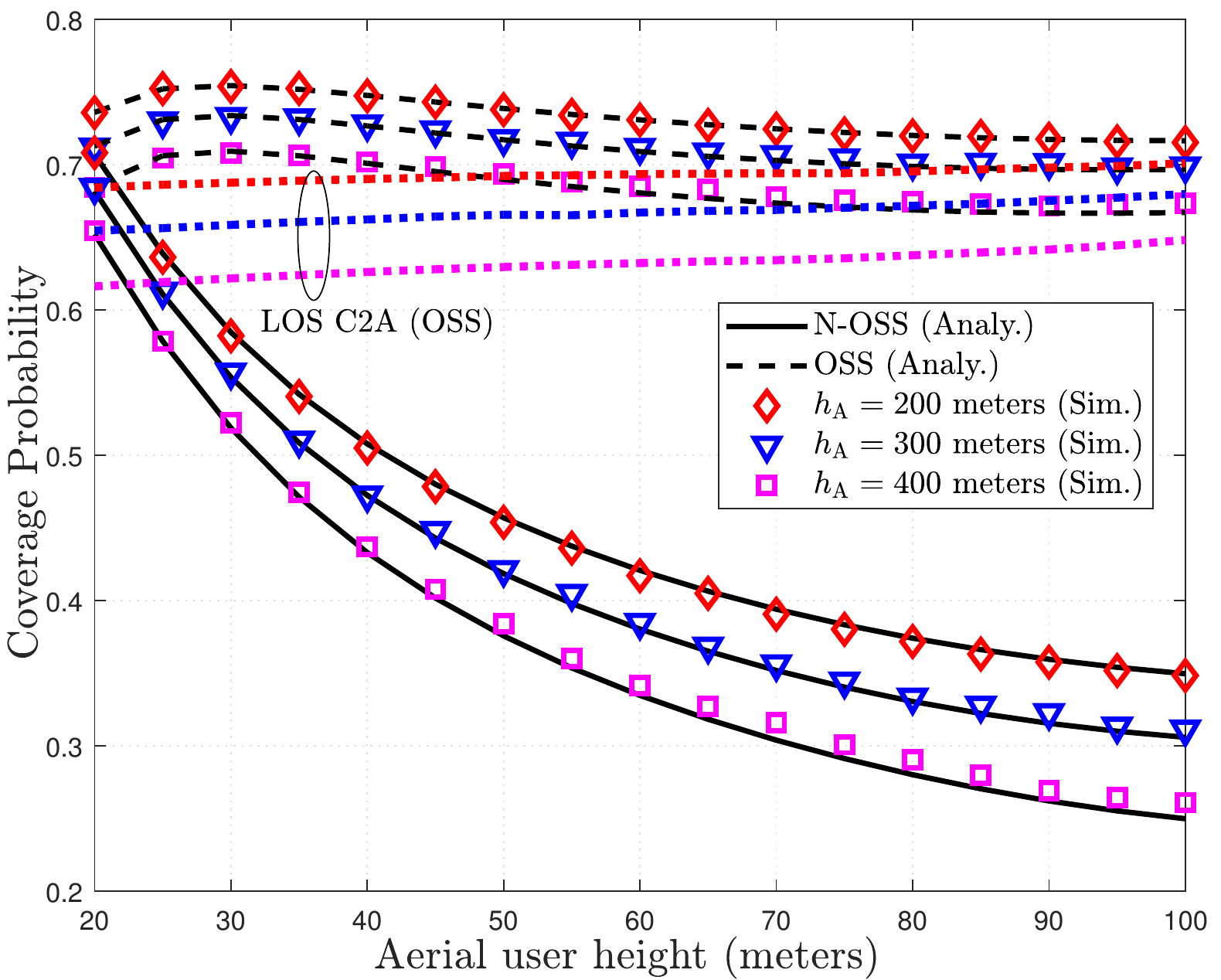}
  \captionof{figure}{Coverage probability versus  aerial user height.}
\label{ht}
\end{minipage}\hfill
\begin{minipage}[t]{.5\textwidth}
  \centering
  \includegraphics[width=1\linewidth]{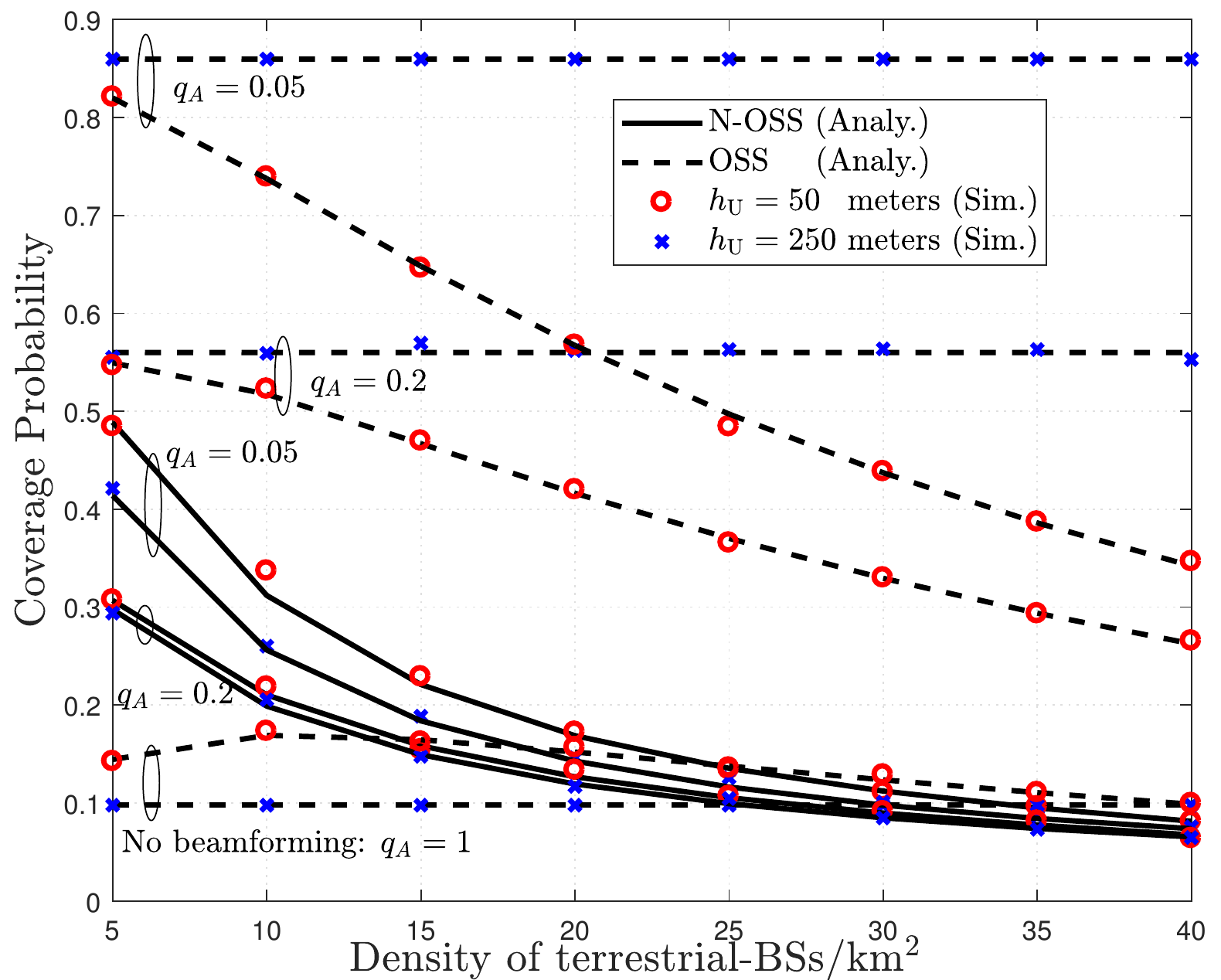}
  \captionof{figure}{Coverage probability versus  terrestrial network density ($ h_{\rm A}=300$ meters).}
\label{density}
\end{minipage}
\end{figure}

In Fig. \ref{ht}, we compare the coverage probability with different aerial user heights with OSS and N-OSS policies. Under the N-OSS policy, as the aerial user height increases, more terrestrial-BSs come within LoS  of the aerial user, which increases the terrestrial interference at a greater rate than the increase in the desired signal power. Under the OSS policy, a different trend can be observed at low aerial user heights, i.e., the coverage probability increases. This is because at very low aerial user heights, the improvement in the desired signal power becomes greater than the increase in the terrestrial interference (since the aerial user may be in LoS with few terrestrial-BSs while terrestrial interference is dominated by NLoS terrestrial-BSs). However, with a further increase in the aerial user height, the interference received from the LoS terrestrial-BSs dominates, which degrades the coverage probability. Indeed, the significant difference in coverage probability under N-OSS and OSS policies is due to the heavy interference that the aerial user experiences under the N-OSS policy which includes aerial and terrestrial interferers alike. Under the OSS policy, by contrast, the aerial user is subject to interference only from BSs that belong to the same network (terrestrial or aerial) as its serving BS. It should also be mentioned that under the assumption that the C2A links are all LoS links, the coverage probability converges to the actual coverage probability. 

Fig. \ref{density} shows the coverage probability versus the density of the terrestrial-BSs for different aerial user heights and  beamwidths under the OSS and N-OSS policies. Under the N-OSS policy, we notice that the coverage probability decreases as the density of terrestrial-BSs increases due to the increase in the terrestrial interference (more LoS terrestrial-BSs exist as the aerial user height increases). We can also observe that for a given density of terrestrial-BSs and regardless of the $q_{\rm A}$, the coverage probability decreases as the aerial user height increases due to the increase in the terrestrial interference (terrestrial interference is dominated by LoS terrestrial-BSs). By contrast, under the OSS policy and at a high aerial user height (250 meters), the density of the terrestrial-BSs has no impact on the coverage probability, since the aerial user is always associated with an aerial-BS according to Fig. \ref{ass}, and thus, it does only receive interference from aerial-BSs. However, a similar trend is not observed if the aerial user is close to the terrestrial-BSs. For instance, at a low aerial user height (50 meters), the coverage probability decreases as the density of the terrestrial-BSs increases, since  the aerial user may be associated with a terrestrial-BS, and so increasing the density of terrestrial-BSs increases the terrestrial interference. Overall, aerial BSs with more directed beams (smaller beamwidth implies smaller $q_{\rm A}$) improves the coverage probability because it decreases the aerial interference power.

\begin{figure}
\captionsetup{width=0.47\textwidth}
\centering
\begin{minipage}[t]{.5\textwidth}
  \centering
  \includegraphics[width=1\linewidth]{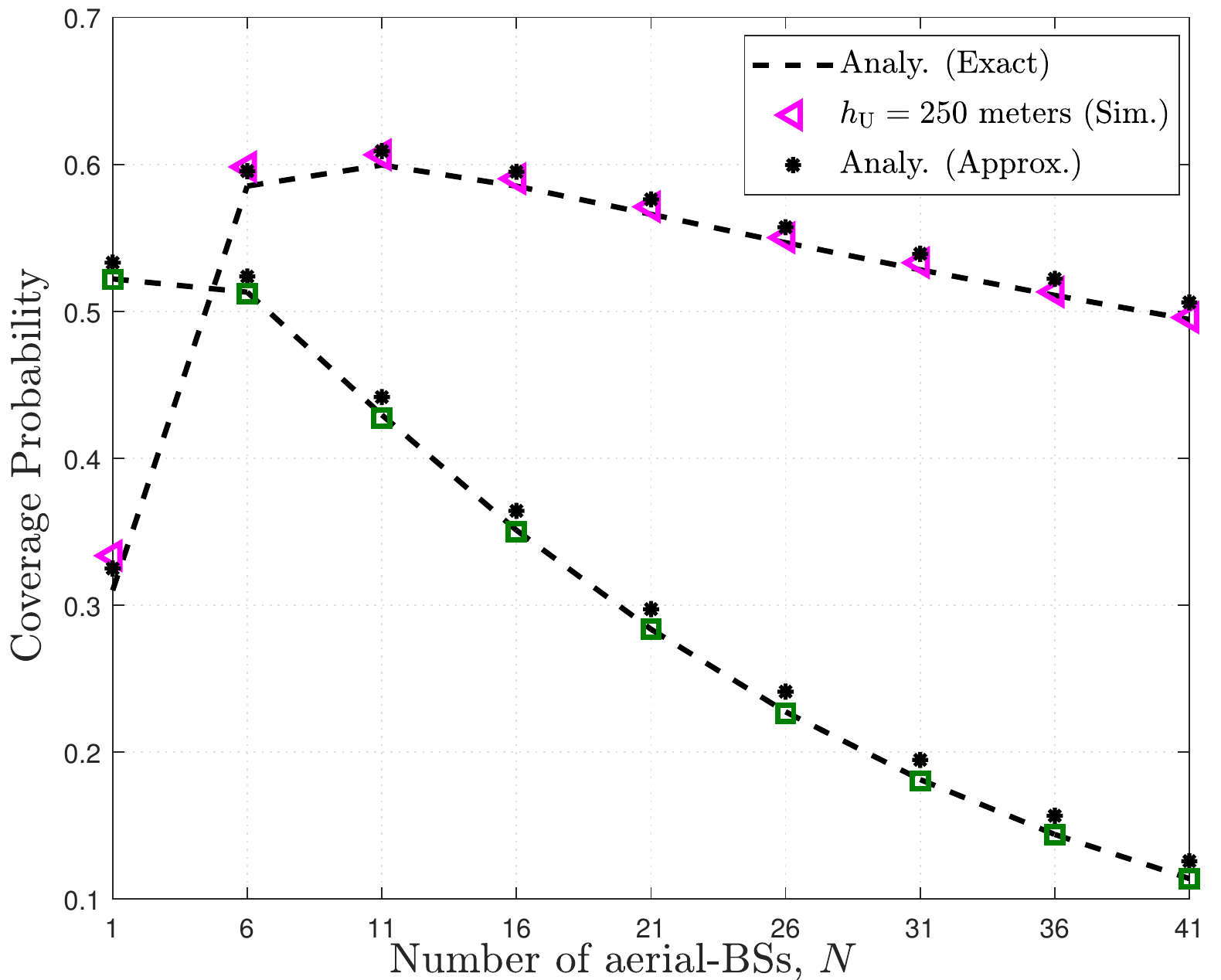}
  \captionof{figure}{Coverage probability versus  number of aerial-BSs with $r_{\rm D}= 1$ km.}
\label{NAB}
\end{minipage}\hfill
\begin{minipage}[t]{.5\textwidth}
  \centering
  \includegraphics[width=1\linewidth]{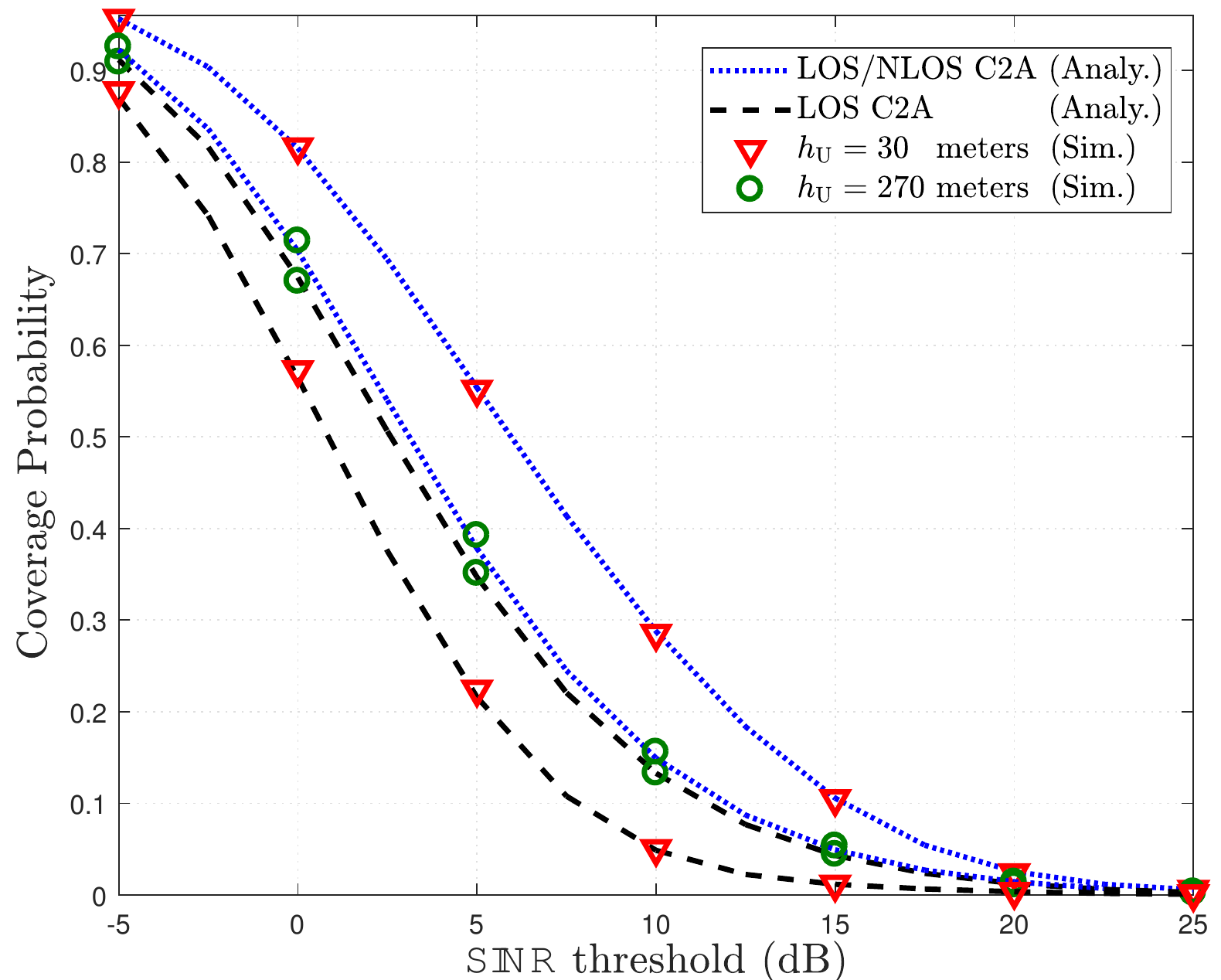}
  \captionof{figure}{Coverage probability versus  \textsf{SINR} threshold with $ h_{\rm A}=300$ meters.}
\label{LOS}
\end{minipage}
\end{figure}

  We plot the coverage probability versus the number of aerial-BSs for different  aerial user heights in Fig. \ref{NAB}. The approximation of the coverage probability using the tight lower bound of the Gamma distribution's CCDF is in a good agreement with the exact performance. Generally, increasing the number of aerial-BSs decreases the coverage probability due to the fact that the increase in the aerial interference is greater than the increase in the desired signal. However, at high aerial user heights (e.g., 250 meters) and low aerial-BS numbers, the coverage probability increases as the number of aerial BSs increases. This is because the increase in aerial interference power is low (small number of aerial-BSs) compared to the improvement in the desired aerial signal. Overall, denser aerial-BSs  decreases the coverage probability especially at low aerial user heights.

The validation of the LoS C2A assumption, which is presented in Section \ref{sub:LOS}, is shown in Fig. \ref{LOS}. At low  aerial user heights (e.g., $h_{\rm U}=30$ meters), LoS C2A assumption is not a good approximation. This is because of the high likelihood of obstructions (NLoS links) between the aerial user and  terrestrial-BSs. However, at elevated heights (e.g., $h_{\rm U}=270$ meters), the aerial user is very likely to be in LoS with  terrestrial-BSs (no or a few NLoS terrestrial-BSs exist) which justifies the correctness and accuracy of the LoS C2A assumption.

\begin{figure}[!t]
	\centering
	\includegraphics[scale=0.53]{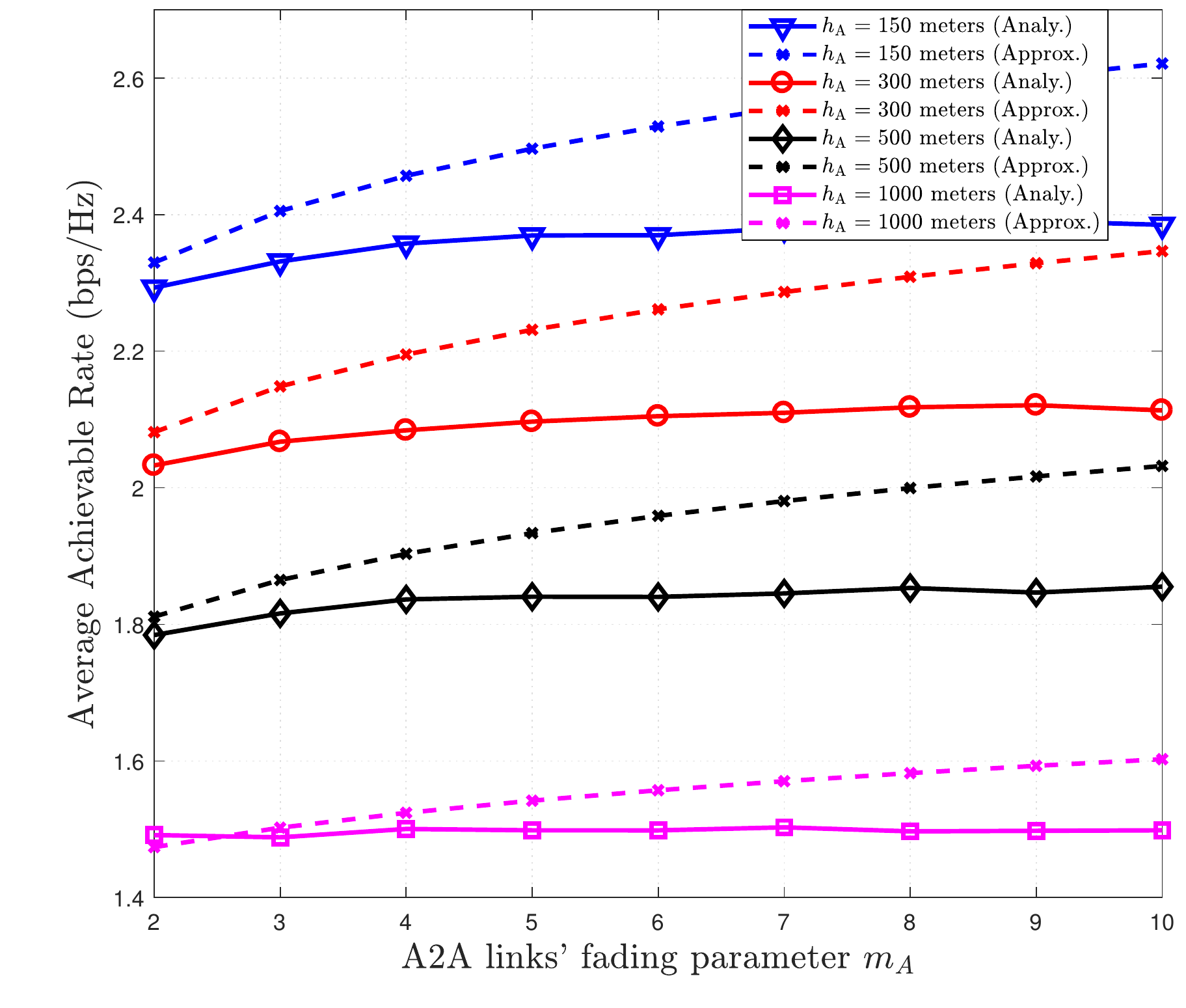}
	\caption{Average achievable rate versus fading parameter  $m_A$ for A2A links, where $ h_{\rm U}=50$ meters.}
	\label{ratemA}
\end{figure}

In Fig. \ref{ratemA}, we show the accuracy of  the average achievable rate approximation given in (\ref{eq:Rnuapp}). NLoS links are assumed to follow Rayleigh fading, $m_N=1$, whereas  LoS links follow Nakagami-$m$ with $m_L=2$. We illustrate the average achievable rate for different fading parameter $m_A$ for the A2A channels. According to the figure, the approximate rate derived in (\ref{eq:Rnuapp}) matches very closely the exact analytical results for a smaller fading parameter $m_{\rm A}$. As the latter increases, the approximation in (\ref{eq:Rnuapp}) becomes an upper bound for the exact average achievable rate. Moreover, as the aerial-BSs height increase, the rate decreases. This can be explained by the fact that the power signal received from the aerial-BSs is weakened by the high path-loss, whereas the aggregate interference power received from the terrestrial-BSs remains the same.

\section{Conclusion}
Using stochastic geometry tools, we proposed a complete framework to analyze the coverage probability and rate of a typical aerial user served by a network of aerial-BSs and LoS/NLoS terrestrial-BSs. Useful distance distributions between the aerial user and its nearest terrestrial-BSs (LoS and NLoS) and aerial-BS were derived, which enabled the computation of the association probabilities of the typical aerial to terrestrial and aerial networks. Exact and approximations of the coverage probability and average achievable rate were presented assuming Nakagami-$m$ fading for all C2A and A2A channels. Simulations revealed that the approximations of the coverage probability and average achievable rate agree very well with the exact expressions, especially with a small fading parameter. Simple expressions were presented under the assumption of only LoS transmissions in the C2A links.
Several conclusions can be drawn from the simulations and evaluated analytical expressions:
\begin{itemize}
\item The association probabilities of an aerial user with an aerial-BS or terrestrial-BS depend on the height of the aerial user, the height of the aerial-BS, and the environment. For instance,  the aerial user tends to connect to an aerial-BS when the former is hovering at a very low height and in a dense environment. It was also shown that an optimum aerial user height exists that maximizes the association probability to a terrestrial-BS. As the aerial user's height increases, the terrestrial-BS is no longer able to provide enough signal power due to high path-loss, which means the aerial user will then certainly be served by an aerial-BS.

\item Simulation results show that the aerial user receives strong interference signals from  LoS terrestrial-BSs and aerial-BSs, which degrade the coverage probability and rate. Orthogonal spectrum sharing between terrestrial and aerial networks considerably increases  the coverage and rate performance of the aerial user. The simulations shows that an optimum height exists for which the coverage probability of the aerial user is maximized.

\item  It was also shown that employing directive beamforming at LoS aerial-BSs is essential in the operation of VHetNets as it decreases the aggregate power of aerial-BSs interference, and as a result it increases the aerial user coverage substantially. Other interference mitigation techniques might also be deployed to improve the aerial user operation.

\item  The results further show that aerial users at elevated heights are mainly in LoS with the terrestrial-BSs (no or a few NLoS links exist). Thus, the analysis can be simplified by assuming two tiers of BSs (i.e., LoS terrestrial-BSs and aerial-BSs). The simulation proves that this approximation is accurate at elevated heights and matches well the performance with LoS/NLoS C2A links. However, this approximation is not valid for low aerial user heights where NLoS links are more likely to exist and cannot be dismissed in the performance evaluation.

\item  Finally, it can be concluded from the results that increasing the density of the aerial-BSs may or may not degrade the performance, depending on the height of both the aerial-BSs and the aerial user. As we saw, a denser aerial network decreases the coverage probability of the aerial user, especially when it hovers at low heights.
\end{itemize}

\appendices
\vspace{-0.6cm}
\renewcommand{\thesection}{\Alph{section}}
\section{Proof of Lemma \ref{lem:1}}
\label{ap:RL}
$R_\nu$ can be written as $R_\nu=\sqrt{Z_\nu^2-h_{\rm UT}^2}$, $\nu \in\{\rm L,N\}$, where $Z_\nu$ is the horizontal distance between the aerial user and the nearest LoS and NLoS terrestrial-BSs. Using the null probability of PPP \cite{andrews}, we have
\vspace{-0.3cm}
\begin{align}
F_{R_\nu}(r)&=1-\mathbb{P}\left(R_\nu\geqslant r\right)\overset{(a)}{=}1-\exp\left(-2\pi\int_{h_{\rm UT}}^{r}\lambda_{\rm T} t P_\nu(t)dt\right).
\end{align}
Since $r=\sqrt{z^2+h_{\rm UT}^2}$, taking the integral in (a) with respect to $z$ completes the proof of (\ref{eq:cdfRNRL}) (note that $t$ is a dummy variable). (\ref{eq:pdfRNRL}) can be obtained by differentiating $F_{R_\nu}(r)$ with respect to $r$.
\vspace{-1cm}
\section{Proof of Lemma \ref{lem:AL}}
\label{ap:Lemma1}
The association probability of an aerial user with an LoS terrestrial-BS can be written as
\begin{eqnarray}
\label{eq:ap1}
 \mathcal{A}_{\rm L}&=&\mathbb{P}\left( \mu_L R_{L}^{-\alpha_{L}} \geqslant \mu_{N} R_{N}^{-\alpha_{N}};\mu_L R_{L}^{-\alpha_{L}} \geqslant \mu_AR_{A}^{-\alpha_{A}}\right)\nonumber\\
&\overset{(a)}{=}&\mathbb{P}\left( \mu_L R_{L}^{-\alpha_{L}}\!\!\geqslant \mu_{N} R_{N}^{-\alpha_{N}}\right)\!\!\times \mathbb{P}\left(\mu_L R_{L}^{-\alpha_{L}}\!\!\geqslant \mu_AR_{A}^{-\alpha_{A}}\right)\nonumber\\
&=&\mathbb{P}\left(\!R_N \geqslant\tau_{N|\mathcal{E}_{\rm L}}(r)\right)\!\!\times \mathbb{P}\left(\!R_A \geqslant \tau_{A|\mathcal{E}_{\rm L}}(r)\right)\label{eq:assdef}\\
&\overset{(b)}{=}&\left(\int_{h_{\rm UT}}^{\infty}F^{(c)}_{R_N}\left( \tau_{N|\mathcal{E}_{\rm L}}(r)\right)f_{R_L(r)} dr\right) \times \left(\int_{h_{\rm UT}}^{\infty}F^{(c)}_{R_A }\left(\tau_{A|\mathcal{E}_{\rm L}}(r)\right)f_{R_L}(r)dr\right),
\end{eqnarray}
where (a) follows from the independence of the two point processes that represent the aerial-BSs and  terrestrial-BSs, (b) follows from  the definition of the Complementary CDF and averaging over $R_L$. Finally, using (\ref{eq:cdfRNRL}) and (\ref{eq:FrA}) along with some mathematical manipulations completes the proof.
\vspace{-1cm}
\section{Proof of Lemma 3}
\label{ap:Lemma3}
The CDF of ${\tilde{R}_L}$ can be written as
\begin{eqnarray}
\label{eq:rlrild}
F_{\tilde{R}_L| \mathcal{E}_{\rm L}}(r)&=& \mathbb{P}\left[R_L<r| \mathcal{E}_{\rm L}\right]\overset{(a)}{=}\frac{\mathbb{P}\left[R_L<r; \mathcal{E}_{\rm L}\right]}{\mathbb{P}[\mathcal{E}_{\rm L}]}\overset{(b)}{=}\frac{\mathbb{P}\left[R_L\leqslant r; (R_A\geqslant \tau_{A|\mathcal{E}_{\rm L}}(r);R_N\geqslant \tau_{N|\mathcal{E}_{\rm L}}(r))\right]}{ \mathcal{A}_{\rm L}}\nonumber\\
&=&\frac{\mathbb{P}\left[R_L\leqslant r; R_A\geqslant \tau_{A|\mathcal{E}_{\rm L}}(r)\right]\mathbb{P}\left[R_L\leqslant r;R_N\geqslant \tau_{N|\mathcal{E}_{\rm L}}(r)\right]}{ \mathcal{A}_{\rm L}}\nonumber\\
&\overset{(c)}{=}&\frac{1}{ \mathcal{A}_{\rm L}}\int_{h_{\rm UT}}^{r}F^{(c)}_{R_A}(\tau_{A|\mathcal{E}_{\rm L}}(r))F^{(c)}_{R_N}(\tau_{N|\mathcal{E}_{\rm L}}(r)) f_{R_L}(x)dx,
\end{eqnarray}
where (a) follows from Bayes' rule, (b) is obtained from (\ref{eq:assdef}), and (c) follows from  averaging over $R_L$. Finally, differentiating (\ref{eq:rlrild}) with respect to $r$ completes the proof.
\vspace{-0.7cm}
\section{Proof of Lemma \ref{lem:LTT}}
\label{ap:LI}
The Laplace transform of the aggregated interference power is given by
\vspace{-0.3cm}
\begin{eqnarray}
\label{eq:LIproof}
&&\mathcal{L}_{(I_N+I_L)|\mathcal{E}_{ \nu} }(s )=\mathbb{E}_{(I_N+I_L)|\mathcal{E}_{ \nu}}\left[\exp(-s  ({I_N+I_L})) \right]\nonumber\\
&&\overset{(a)}{=}\mathbb{E}_{\Phi_{\rm L}}\left[\!\prod_{x_j\in\Phi_{\rm L} \setminus x_0}^{}\!\!\mathbb{E}_{{\rm H}_{\rm L}^{x_j}}\!\!\left[\exp\left(\!-s\mu_L {\rm H}_{\rm L}^{x_j} d_{L,x_j}^{-\alpha_L}\!\right)\right]\!\right]\!\!\times\!\mathbb{E}_{\Phi_{\rm N}}\!\!\left[\!\prod_{x_j\in\Phi_{\rm N} \setminus x_0}^{}\!\!\!\mathbb{E}_{{\rm H}_{\rm N}^{x_j}}\left[\exp\left(\!-s\mu_N {\rm H}_{\rm N}^{x_j} d_{N,x_j}^{-\alpha_N}\right)\right]\!\right]\nonumber\\
&&\!\overset{(b)}{=}\prod_{\omega\in\{\rm L,N\}}^{}\!\!\!\left[\!\mathbb{E}_{\Phi_{\rm \omega}}\!\!\left[\prod_{x_j\in\Phi_\omega \setminus x_0}^{}\left(\frac{m_\omega}{m_\omega+s\mu_\omega d_{\omega,x_j}^{-\alpha_\omega}}\right)^{m_\omega}\!\right]\!\right],
\end{eqnarray}
where (a) follows from  (\ref{eq:int}) and the independence of the small-scale fading and PPP, and (b) is obtained from the moment generating function of the Gamma distribution. Finally, using the probability generating functional (PGFL)  and the results in Table \ref{my-label3} complete the proof.
\vspace{-0.7cm}
\section{Proof of Lemma \ref{lem:LTA}}
\label{ap:LIA}
It was proven in \cite[Corollary 3]{chetlur} that when the serving BS (either terrestrial-BS or aerial-BS) is located at a distance $r$ from the aerial user, the distribution of the distance between the aerial user and the j-th interfering aerial-BS $d_{A,x_j}, j \in [1,N']$ is given by
\begin{equation}
\label{eq:ui}
f_{d_{A,x_j}}(d_j)=\left\{
\begin{array}{ll}
\frac{2 d_j}{d^2-r^2},& r\leqslant d_j\leqslant d\\
0, &\text{Otherwise },
\end{array}
\right.
\end{equation}
where $N'$ is the number of interfering aerial-BSs. It should be noted that there are $N'-i$ interfering aerial-BSs that have a gain of $G_{\rm m}^{\rm A}$  with a probability $q_{\rm A}$ and $i$ interfering aerial-BSs that have a gain of $g_{\rm s}^{\rm A}$ with a probability of $1-q_{\rm A}$.  Hence,  $i$ follows a Binomial distribution $\mathcal{B}(N',q_{\rm A})$. Indeed, the two subsets of interfering aerial-BSs (those with a gain of $G_{\rm m}^{\rm A}$ and those with a gain $g_{\rm s}^{\rm A}$) are dependent where the joint distribution of $N'-i$ and $i$ is a multinomial distribution on $N'$ trials with success probabilities $q_{\rm A}$ and $1-q_{\rm A}$, respectively.

The Laplace transform of the interference received from aerial-BSs is given by
\begin{eqnarray}
\label{eq:LIproofA}
\!\!\!\!\!\!\!\!\!\!\!\!\!\!\!&&\!\!\!\!\!\!\!\!\!\!\mathcal{L}_{I_A|\mathcal{E}_{ \nu} }(s )=\mathbb{E}_{I_A|\mathcal{E}_{ \nu} }\left[\exp(-s  {I_A}) \right]=\mathbb{E}_{d_{A,x_j}}\left[\prod_{j=1}^{N'}\mathbb{E}_{{\rm H}^{x_j}_{A}}\left[\exp\left(-sP_A G^{\rm A}(d_j) {\rm H}^{x_j}_{A} d_{j}^{-\alpha_A}\right)\right]\right]\nonumber\\
&\overset{(a)}{=}&\!\!\prod_{j=1}^{N'}\left[\mathbb{E}_{d_{A,x_j}}\!\!\left[\left(\frac{m_A}{m_A+sP_A G^{\rm A}(d_j) d_j^{-\alpha_A}}\right)^{m_A}\!\right]\right]\overset{(b)}{=}f_{i}(i)\left[\mathbb{E}_{d_{A,x_j}}\left[\left(\frac{m_A}{m_A+sP_A G_{\rm m}^{\rm A} d_j^{-\alpha_A}}\right)^{m_A}\right]\right]^{\!N'\!-i}\nonumber\\
&&\times \left[\mathbb{E}_{d_{A,x_j}}\left[\left(\frac{m_A}{m_A+sP_A g_{\rm s}^{\rm A} d_j^{-\alpha_A}}\right)^{m_A}\right]\right]^{i}\ \nonumber\\
&\overset{(c)}{=}&\sum_{i=0}^{N'}\binom{N'}{i}q_{\rm A}^{N'-i}(1-q_{\rm A})^i\left[\int_{\tau_{A|\mathcal{E}_{\rm \nu}}(r))}^{d}\left(\frac{m_A}{m_A+sP_A G^{\rm A}_{\rm m}\eta_{A}t^{-\alpha_{A}}}\right)^{m_A}\frac{2t}{d^2-r^2}dt\right]^{N'-i}\nonumber\\
&&\times\left[\int_{\tau_{A|\mathcal{E}_{\rm \nu}}(r))}^{d}\left(\frac{m_A}{m_A+s P_A g^{\rm A}_{\rm s}\eta_{A}t^{-\alpha_{A}}}\right)^{m_A}\frac{2t}{d^2-r^2}dt\right]^{i},
\end{eqnarray}
where (a) follows from using the Gamma distribution of the small scale fading channel gain of the A2A link. (b) follows from the binomial distribution of $i$ with $f_i(i)=\sum_{i=0}^{N'}\binom{N'}{i}q_{\rm A}^{N'-i}(1-q_{\rm A})^i$. (c) is obtained after averaging over $d_{j}$. Using the identity $(1+z)^a=1/\Gamma(-a){\rm G}_{1,1}^{1,1}[z\vert  {1-a \atop 0}]$ and \cite[eq(7.811.2)]{grad} along with some additional mathematical manipulations, (\ref{eq:Li1}) is obtained.
\vspace{-0.7cm}

\section{Proof of Theorem \ref{th:cov}}
\label{ap:cov}
The coverage probability conditioned on the event that the aerial user is connected to a BS from $\Phi_\nu$, $\nu \in \{\rm L,N,A\}$, is given by 
{\allowdisplaybreaks
\begin{eqnarray}
\label{eq:Pu}
\!\!\!\!\!\!\!\!\!\!\!\!\!\!\!&&\!\!\!\!\!\!\!\!\!\!\!\!\!\!\!\mathcal{C}_\nu=\mathbb{P}(\gamma\geqslant T)\overset{(a)}{=}\mathbb{E}_{{\rm H}_{\nu}^{x_0},\tilde{R}_\nu,I}\left[\mathbb{P}\left(\frac{\mu_\nu{\rm H}_{\nu}^{x_0}r_{\nu}^{-\alpha_{L}}}{I+\sigma^2}\geqslant T\right)\right]\overset{(b)}{=}\!\mathbb{E}_{\tilde{R}_\nu,V}\left[\mathbb{E}_{{\rm H}_{\nu}^{x_0}}\left[\mathbb{P}\left({\rm H}_{\nu}^{x_0} \geqslant \frac{ r_{\nu}^{\alpha_{L}}T V}{\mu_\nu}\right)\right]\right]\nonumber\\
&\overset{(c)}{=}&\!\!\!\!\mathbb{E}_{\tilde{R}_\nu}\!\!\left[\!\mathbb{E}_{V}\!\!\left[\frac{\Gamma\left(m_\nu,\frac{m_\nu T r^{\alpha_\nu}V}{\mu_\nu}\right)}{\Gamma(m_\nu)}\right]\!\right]\!\!\overset{(d)}{=}\!\!\sum_{k=0}^{m_\nu-1}\!\!\!\frac{1}{k!}\!\!\left(\!\frac{m_\nu T }{\mu_\nu}\!\right)^{\!\!k}\!\mathbb{E}_{\tilde{R}_\nu}\!\!\left[r^{k\alpha_\nu}\mathbb{E}_V\!\left[V^k \exp\left(\!\frac{\!-m_\nu T r^{\alpha_\nu}V}{\mu_\nu}\!\right)\!\right]\!\right],
\end{eqnarray}}
where (a) follows from averaging the coverage probability over the random variables, $\{{\rm H}_{\nu}^{x_0},\tilde{R}_\nu,V\}$ and (b) is obtained by exploiting the independence between the three random variables. Moreover in (\ref{eq:Pu}), (c) follows after applying the CCDF of Gamma-distributed channel gain ${\rm H}_{\nu}^{x_0}$, and (d) is obtained by assuming $m_\nu$ to be an integer and using the series expansion of the upper incomplete Gamma function. Finally, using the identity, i.e., $\mathbb{E}_V\left[V^k \exp(-s V)\right]=(-1)^k\frac{\partial \mathcal{L}_V(s)}{\partial s^k}$, and averaging over $\tilde{R}_\nu$, we obtain (\ref{eq:Pnu}).

\vspace{-0.6cm}

\bibliographystyle{IEEEtran}  
\bibliography{references}  

\end{document}